\documentclass[showpacs,preprintnumbers,amsmath,amssymb]{revtex4}
\usepackage{graphicx}% Include figure files
\usepackage{dcolumn}% Align table columns on decimal point
\usepackage{bm}% bold math

\begin{document}
\def\d{{\rm d}}
%%%%%%%%%%%%%%%%%%%%%%%
\def\Epos{E_{\rm pos}}
\def\ap{\approx}
\def\eff{{\rm eff}}
\def\L{{\cal L}}
\newcommand{\vev}[1]{\langle {#1}\rangle}
\newcommand{\CL}   {C.L.}

\newcommand{\be}{\begin{equation}}
\newcommand{\ee}{\end{equation}}
\newcommand{\ese}{\end{subequations}}
\newcommand{\ba}{\begin{eqnarray}}
\newcommand{\ea}{\end{eqnarray}}
\newcommand{\nbb}{$\beta\beta_{0\nu}$ }
\def\VEV#1{\left\langle #1\right\rangle}
\let\vev\VEV
\def\e6{E(6)}
\def\10{SO(10)}
\def\21{SU(2) $\otimes$ U(1) }
\def\321{$\mathrm{SU(3) \otimes SU(2) \otimes U(1)}$ }
\def\lr{SU(2)$_L \otimes$ SU(2)$_R \otimes$ U(1)}
\def\422{SU(4) $\otimes$ SU(2) $\otimes$ SU(2)}

\setlength{\topmargin}{-5mm}

\newcommand{\Tehran}{%
Institute for Research in Fundamental Sciences (IPM), P.O. Box
19395-5531, Tehran, Iran}
\def\roughly#1{\mathrel{\raise.3ex\hbox{$#1$\kern-.75em
      \lower1ex\hbox{$\sim$}}}} \def\lsim{\roughly<}
\def\gsim{\roughly>}
\def\ltap{\raisebox{-.4ex}{\rlap{$\sim$}} \raisebox{.4ex}{$<$}}
\def\gtap{\raisebox{-.4ex}{\rlap{$\sim$}} \raisebox{.4ex}{$>$}}
\def\lsim{\raise0.3ex\hbox{$\;<$\kern-0.75em\raise-1.1ex\hbox{$\sim\;$}}}
\def\gsim{\raise0.3ex\hbox{$\;>$\kern-0.75em\raise-1.1ex\hbox{$\sim\;$}}}

\preprint{ IPM/P-2008/053}

%---------------------------------------------------

\title{A Window on the CP-violating Phases of MSSM from Lepton Flavor Violating Processes}

%----------------------------------------------------
\date{\today}
\author{S. Yaser Ayazi}\email{yaserayazi@mail.ipm.ir}
\author{Yasaman Farzan}\email{yasaman@theory.ipm.ac.ir}
\affiliation{\Tehran}

%--------------------------------------------------
\begin{abstract}
It has  recently been shown that by measuring the transverse
polarization of the final particles in the LFV processes $\mu \to
e\gamma$, $\mu \to eee$ and $\mu N\to e N$, one can derive
information on the CP-violating phases of the underlying theory. We
derive formulas for the transverse polarization of the final
particles in terms of the couplings of the effective potential
leading to these processes. We then study the dependence of the
polarizations of $e$ and $\gamma$ in the $\mu \to e \gamma$ and $\mu
N \to e N$ on the parameters of the Minimal Supersymmetric Standard
Model (MSSM). We show that combining the information on various
observables in the $\mu \to e\gamma$ and $\mu N\to e N$ search
experiments with the information on the electric dipole moment of
the electron can help us to solve the degeneracies in parameter
space  and to determine the values of certain phases.

\end{abstract}
%--------------------------------------------------------
%--------------------------------------------------------
\date{\today}
\maketitle
%-------------------------------------------------------------------

%% \bibliographystyle{h-physrev4}
%% \bibliography{lgenesis-ref,valle-ref}

\section{Introduction}
In the framework of the Standard Model (SM), Lepton Flavor
Violating (LFV) processes such as $\mu^+ \to e^+ \gamma$,
$\mu^+\to e^+ e^- e^+$ and $\mu-e$ conversion on nuclei ({\it
i.e.,} $\mu N \to e N$) are forbidden. Within the SM augmented
with neutrino mass and mixing, such processes are in principle
allowed but the rates are suppressed by  factors of $(\Delta
m_\nu^2/E_{W}^2)^2$ \cite{petcov} and are too small to be probed
in the foreseeable future.

Various models beyond the SM can  give rise to LFV rare decay with
branching ratios exceeding the present bounds \cite{pdg}:
$${\rm Br}(\mu^+ \to e^+ \gamma)<1.2 \times 10^{-11} \ \ \ \ {\rm
Br}(\mu^+ \to e^+ e^+e^-)<1.0\times 10^{-12} \ \ {\rm at} \  90\%
\ {\rm C.L.}    $$ For low scale MSSM ($m_{SUSY}\sim 100~{\rm
GeV}$), these experimental bounds imply stringent bounds on the
LFV sources in the Lagrangian.
 The MEG experiment at PSI \cite{MEGhomepage}, which is expected
 to release data in summer 2009, will eventually be able to probe Br$(\mu^+ \to e^+
\gamma)$ down to $10^{-13}$.  In our opinion, it is likely that
the first evidence for physics beyond the SM comes from the MEG
experiment. If the branching ratio is close to its present bound,
the MEG experiment will detect statistically significant number of
such events. As a result, making precision measurement will become
a possibility within a few years. Muons in the MEG experiment are
produced by decay of the stopped pions (at rest) so they are
almost 100\% polarized. This opens up the possibility of learning
about  the chiral nature of the underlying theory by studying the
angular distribution of the final particles relative to the spin
of the parent particle ~\cite{review}. In Ref. \cite{asli}, it has
been shown that by measuring the polarization of the final states
in the decay modes $\mu^+\to e^+ \gamma$ and $\mu^+\to e^+e^-e^+$,
one can derive information on the CP-violating sources of the
underlying theory. Notice that even for the state-of-the-art LHC
experiment, it will be quite challenging (if possible at all) to
determine the CP-violating phases in the lepton sector
\cite{godbole}. Suppose the LHC establishes a particular theory
beyond the SM such as supersymmetry. In order to learn more about
the CP-violating phases, the well-accepted strategy is to build
yet a more advanced accelerator such as ILC. Considering the
expenses and challenges before constructing such an accelerator,
it is worth to give any alternative method such as the one
suggested in Ref. \cite{asli} a thorough consideration. In this
paper we elaborate more on this method within the framework of
R-parity conserving MSSM.

 LFV
sources in the Lagrangian can also give rise to sizeable $\mu-e$
conversion rate. There are strong bounds on the rates of such
processes \cite{sindrum,review,Wintz}: \be R (\mu  {\rm Ti}\to e
{\rm Ti})\equiv {\Gamma(\mu {\rm Ti} \to e {\rm Ti}) \over
\Gamma(\mu {\rm Ti} \to {\rm capture})}<6.1 \times 10^{-13}\ \ \ .
\label{RRR}\ee The upper bound on $R$ restricts the LFV  sources
however, for the time being, the bound from $\mu \to e\gamma$ is
more stringent. The PRISM/PRIME experiment is going to perform a
new search for the $\mu-e$ conversion \cite{Prism}. In case that
the values of LFV parameters are close to the present upper bound,
a significantly large number of the $\mu-e$ conversion events can
be recorded by PRISM/PRIME. Recently it is shown in \cite{Sacha}
that if the initial muon is polarized (at least partially),
studying the transverse polarization of the electron  yields
information on the CP-violating phase. In this paper, we elaborate
more on this possibility taking into account all the  relevant
effects in the context of R-parity conserving MSSM.

In the end of the paper, we study the possibility of eliminating
the degeneracies of the parameter space by combining information
from $\mu \to e \gamma$ and $\mu-e$ conversion experiments. We
then demonstrate   that the forthcoming results from $d_e$ search
can help us to eliminate the degeneracies further (cf.
Figs.~(\ref{degeneracy}-a,\ref{degeneracy}-b)).

The paper is organized as follows:
 In sec.
\ref{polarization}, using the results of Ref. \cite{asli}, we
calculate the polarization of the final particles in decay $\mu
\to e \gamma$  in terms of the couplings of the low energy
effective Lagrangian (after integrating out the supersymmetric
states). We also briefly discuss $\mu \to e e e$ and the
challenges of deriving the CP-violating phases by its study. In
sec.
 \ref{conversion}, we calculate the transverse polarization of the
 muon in the $\mu-e$ conversion experiment in terms of the
 couplings in the effective Lagrangian which give the dominant
 contribution to $\mu N \to e N$ within the MSSM. In sec.
 \ref{impact},
  we study the overall pattern of the variation
 of $\overline{\langle s_{T_2}\rangle}$ and $\overline{\langle
 P_{T_1}s_{T_2}\rangle}$ with phases and discuss the regions of
 the parameter space where the sensitivity to the phases are
 sizeable. In sec. \ref{degeneracies}, we discuss how by
 combining
 information from the $\mu \to e \gamma$ and $\mu N \to e N$
 experiments, we can solve the degeneracies in the parameter
 space. The conclusions are summarized in sec.~\ref{results}.
%%%%%%%%%%%%%%%%%%%%%%%%%%%%%%%%%%%%%%%%
%%%%%%%%%%%%%%%%%%%%%%%%%%%%%%%%%%%%%%%%%555
\section{Polarization of the final particles \label{polarization}}
%%%%%%%%%%%%%%%%%%%%%%%%%%%%%%%%%%%%%%%%%%%55
%%%%%%%%%%%%%%%%%%%%%%%%%%%%%%%%%%%%%%%%5555
The low energy effective Lagrangian that gives rise to $\mu \to e
\gamma$ can be written as \be {\cal L}=\frac{A_R}{m_\mu}
\bar{\mu}_R \sigma^{\mu \nu}e_L F_{\mu \nu}+\frac{A_L}{m_\mu}
\bar{\mu}_L \sigma^{\mu \nu} e_R F_{\mu \nu}+\frac{A_R^*}{m_\mu}
\bar{e}_L \sigma^{\mu \nu}\mu_R F_{\mu \nu}+\frac{A_L^*}{m_\mu}
\bar{e}_R \sigma^{\mu \nu} \mu_L F_{\mu \nu} \ ,
\label{LFV-lagrangian}\ee where $\sigma^{\mu
\nu}=\frac{i}{2}[\gamma^\mu,\gamma^\nu]$ and $F_{\mu \nu}$ is the
photon field strength: $F_{\mu \nu}=\partial_\mu \varepsilon_\nu
-\partial_\nu \varepsilon_\mu$. $A_L$ and $A_R$ receive
contributions from the LFV parameters of MSSM at one loop level
\cite{hisano,okada,review}. In this section we derive the
polarizations of the final particles in the LFV rare decays in
terms of $A_L$ and $A_R$. Let us define the longitudinal and
transverse directions as follows:
$\hat{l}\equiv\vec{p}_{e^+}/|\vec{p}_{e^+}|$, $\hat{T_2}\equiv
\vec{p}_{e^+}\times {\vec{s}_{\mu}}/|\vec{p}_{e^+}\times
{\vec{s}_{\mu}}|$ and $\hat{T_1}\equiv\hat{T_2}\times \hat{l}$.
As shown in \cite{asli}, the partial decay rate of an anti-muon at
rest into a positron and a photon with definite spins of
$\vec{s}_e$ and $\vec{s}_\gamma$ is
 \be \frac{d \Gamma [ \mu^+(P_{\mu^+})\to e^+(P_{e^+},
\vec{s}_{e^+}) \gamma(P_\gamma, \vec{s}_\gamma)]}{d \cos
\theta}=\frac{ m_\mu}{8 \pi}
\left[|\alpha_+|^2|A_L|^2(1+\mathbb{P}_\mu \cos \theta) \sin^2
\frac {\theta_s}{2}+\right.\label{no-averaging}\ee
$$ \left. |\alpha_-|^2|A_R|^2(1-\mathbb{P}_\mu \cos \theta)\cos^2
\frac{\theta_s}{2} - \mathbb{P}_\mu \textrm{Re}[\alpha_+\alpha_-^*
A_L^* A_R e^{i \phi_s} ] \sin \theta \sin \theta_s\right],
$$ where $\mathbb{P}_\mu$ is the polarization of the anti-muon,
$\theta$ is the angle between the directions of the spin of the
anti-muon and the momentum of the positron, and   $\theta_s$ is
the angle between the spin of the positron and its momentum. In
the above formula, $\phi_s$ is the azimuthal angle that the spin
of the final positron makes with the plane of spin of the muon and
the momentum of the positron. Finally, $\alpha_+$ and $\alpha_-$
give the polarization of the final photon: $$
\vec{\varepsilon}\cdot \hat{T}_1\equiv\sum_{j\in \{1,2,3\}}
(\hat{T_1})_j\varepsilon_j ={\alpha_++\alpha_- \over \sqrt{2}} \ \
{\rm and} \ \ \vec{\varepsilon}\cdot \hat{T}_2 \equiv \sum_{j \in
\{1,2,3\}}(\hat{T_2})_j \varepsilon_j ={\alpha_+-\alpha_- \over
\sqrt{2}}i$$ where $\sqrt{|\alpha_+|^2+|\alpha_-|^2}=1$. Notice
that for  a given polarization of the positron, the photon has a
definite polarization: {\it i.e.,} setting $\mathbb{P}_\mu=100\%$
and $\alpha_+=\alpha_- e^{-i\phi_s} (A_R^*/A_L^*)\tan \theta/2
\cot \theta_s/2$, we find $d\Gamma/d\cos \theta=0$. Consider the
case that $\mathbb{P}_\mu=100\%$ and the positron is emitted in
the direction of the spin of the muon; {\it i.e.,} $\theta=0$.
From (\ref{no-averaging}), we find that  for $\theta_s=\pi$ and
$\alpha_+=1$, $d\Gamma/d\cos \theta$ is maximal. In other words,
in this case, the spins of the positron and the photon are
respectively aligned in the  direction anti-parallel and parallel
to the spin of the muon. This is expected because when $\theta=0$
there is a cylindrical symmetry around the axis parallel to the
spin of the muon and therefore the total angular momentum in the
direction of the spin does not receive any contribution from the
relative angular momentum. This means the sum of spins in the
$\hat{l}$ direction has to be conserved which in turn implies that
the decay rate is maximal at $\theta_s=\pi$ and $\alpha_+=1$.
Similar consideration also applies to the case that the positron
is emitted antiparallel to the spin of the muon: For $\theta=\pi$,
the emission is maximal at $\theta_s=0$ and $\alpha_-=1$.

Summing over the polarization of the final particles in
Eq.~(\ref{no-averaging}), we obtain
$$\sum_{\vec{s}_{\gamma}\vec{s}_{e^+}} \frac{d \Gamma [ \mu^+(P_{\mu^+})\to e^+(P_{e^+},
\vec{s}_{e^+}) \gamma(P_\gamma, \vec{s}_\gamma)]}{d \cos
\theta}=\frac{ m_\mu}{8 \pi} \left[|A_L|^2(1+\mathbb{P}_\mu \cos
\theta) + |A_R|^2(1-\mathbb{P}_\mu \cos \theta)\right].
$$ Thus, $\Gamma(\mu \to e \gamma)$ is given by $(|A_L|^2+|A_R|^2)$. It is
convenient to define \be \label{R1} R_1 \equiv {|A_L|^2-|A_R|^2
\over |A_L|^2+|A_R|^2} \ . \ee By measuring the total decay rate
and the angular distribution of the final particles, one can
derive absolute values $A_L$ and $A_R$. To measure the relative
phase of these couplings, the polarization of the final particles
also have to be measured.

Let us define the polarizations of the electron and photon in an
arbitrary direction $\hat{T}$ respectively as \be \langle
s_{T}\rangle\equiv {\sum_{\vec{s}_\gamma}\left[
d\Gamma\left[\mu^+\to
e^+(\vec{s}_{e^+}=\frac{1}{2}\hat{T})\gamma(\vec{s}_\gamma)\right]-
d\Gamma\left[\mu^+\to
e^+(\vec{s}_{e^+}=-\frac{1}{2}\hat{T})\gamma(\vec{s}_\gamma)\right]\right]
\over \sum_{\vec{s}_\gamma\vec{s}_{e^+}}d\Gamma\left[\mu^+\to
e^+(\vec{s}_{e^+})\gamma(\vec{s}_\gamma)\right]} \ee
 %%%%%%%%%%%%%%%%%%%%%%%%%%%%%%%%%%%%%%%
and
%%%%%%%%%%%%%%%%%%%%%%%%%%%%%%%%%%%%5
 \be \langle P_{T}\rangle\equiv {\sum_{\vec{s}_{e^+}}
d\Gamma\left[\mu^+\to
e^+(\vec{s}_{e^+})\gamma(\vec{\varepsilon}\shortparallel\hat{T})\right]
\over \sum_{\vec{s}_\gamma\vec{s}_{e^+}}d\Gamma\left[\mu^+\to
e^+(\vec{s}_{e^+})\gamma(\vec{s}_\gamma)\right]} \ee where
$\vec{\varepsilon}$ is the polarization vector of the photon.

 From Eq.~(\ref{no-averaging}),
we find that the polarization of positron (once we average over
the polarizations of the photon) is
$$\langle s_{T_1}\rangle=\langle s_{T_2}\rangle =0 \ , \ \ \langle
s_l\rangle=\frac{|A_R|^2(1-\mathbb{P}_\mu
\cos\theta)-|A_L|^2(1+\mathbb{P}_\mu\cos\theta)}{|A_R|^2(1-\mathbb{P}_\mu
\cos\theta)+|A_L|^2(1+\mathbb{P}_\mu\cos\theta)}\ .$$ That is
while the linear polarization of the photon (once we sum over the
polarization of the positron) is
$$\langle P_{T_1}\rangle=\langle P_{T_2}\rangle=\frac{1}{2}.$$

Unfortunately, neither the polarization of the positron nor the
polarization of the photon carries any information on the relative
phase of $A_L$ and $A_R$. However, the double correlation of the
polarization carries such information. Let us define double
correlation as follows \be \langle P_{T'} s_{T} \rangle\equiv
{d\Gamma\left[\mu^+\to
e^+(\vec{s}_{e^+}=\frac{1}{2}\hat{T})\gamma(\vec{\varepsilon}\shortparallel
\hat{T'})\right]- d\Gamma\left[\mu^+\to
e^+(\vec{s}_{e^+}=-\frac{1}{2}\hat{T})\gamma(\vec{\varepsilon}\shortparallel\hat{T'})\right]
\over \sum_{\vec{s}_\gamma\vec{s}_{e^+}}d\Gamma\left[\mu^+\to
e^+(\vec{s}_{e^+})\gamma(\vec{s}_\gamma)\right]} \ee where
$\hat{T}$ and $\hat{T'}$ are arbitrary directions. From
Eq.~(\ref{no-averaging}), we find
 \be \label{one-theta}\langle P_{T_1} s_{T_1}\rangle
=-\langle P_{T_2} s_{T_1} \rangle ={- \mathbb{P}_\mu {\rm
Re}[A_L^* A_R] \sin \theta \over |A_R|^2 (1-\mathbb{P}_\mu \cos
\theta)+|A_L|^2 (1+\mathbb{P}_\mu \cos \theta)} \ee and \be
\label{yek-theta}\langle P_{T_1} s_{T_2}\rangle =-\langle P_{T_2}
s_{T_2} \rangle ={ \mathbb{P}_\mu {\rm Im}[A_L^* A_R] \sin \theta
\over |A_R|^2 (1-\mathbb{P}_\mu \cos \theta)+|A_L|^2
(1+\mathbb{P}_\mu \cos \theta)}. \ee Thus, as pointed out in
\cite{asli}, to extract the CP-violating phases both polarization
and their correlation have to be measured. Eq.~(\ref{one-theta})
gives the correlation of the polarizations for particles emitted
along the direction described by $\theta$. Averaging over
$\theta$, we find \be \label{average-theta}\overline{\langle
P_{T_1} s_{T_1}\rangle} =-\overline{\langle P_{T_2} s_{T_1}
\rangle} ={\int_{-1}^1 \mathbb{P}_\mu {\rm Re}[A_L^* A_R] \sin
\theta d\cos \theta\over \int_{-1}^1\left[ |A_R|^2
(1-\mathbb{P}_\mu \cos \theta)+|A_L|^2 (1+\mathbb{P}_\mu \cos
\theta)\right] d\cos \theta}={-\pi\mathbb{P}_\mu {\rm Re}[A_L^*
A_R] \over 4(|A_L|^2+|A_R|^2)} \ee and \be \label{average-theta2}
\overline{\langle P_{T_1} s_{T_2}\rangle} =-\overline{\langle
P_{T_2} s_{T_2} \rangle} ={\int_{-1}^1  \mathbb{P}_\mu {\rm
Im}[A_L^* A_R] \sin \theta d\cos \theta \over\int_{-1}^1\left[
|A_R|^2 (1-\mathbb{P}_\mu \cos \theta)+|A_L|^2 (1+\mathbb{P}_\mu
\cos \theta)\right]d\cos \theta}={\pi\mathbb{P}_\mu {\rm Im}[A_L^*
A_R] \over 4(|A_L|^2+|A_R|^2)}. \ee Notice that to take average
over angles, one should weigh the polarization of positron emitted
within a given interval $(\theta,\theta+d\theta)$ with the number
of emission in this interval and then integrate over angles. That
is why we have integrated over $d \cos \theta$ in both the
numerator and denominator of the right-hand side of the ratios in
Eqs. (\ref{one-theta},\ref{yek-theta}) instead of calculating
$\int \langle P_{T_i}s_{T_j} \rangle d \cos \theta/\int d \cos
\theta$.

From Eqs. (\ref{one-theta},\ref{yek-theta}), we find that if the
polarimeter is located at $\theta=\pi/2$, the polarization and
therefore sensitivity is maximal. Notice that $$\langle
P_{T_i}s_{T_j}\rangle |_{\theta={\pi\over 2}}=\frac{4}{\pi}
\overline{\langle P_{T_i}s_{T_j}\rangle}\ . $$ Measurement of
$\overline{\langle P_{T_i}s_{T_j}\rangle}$ requires setting
polarimeters all around the region where the decay takes place. In
sec. \ref{impact}, we perform an analysis of $\overline{\langle
P_{T_i}s_{T_j}\rangle}$. Up to a factor of $4/\pi$, our results
applies to the case that measurement of the polarization is
performed only at $\theta=\pi/2$.

The  ratios of the polarizations yield the relative phase of the
effective couplings
$$\frac{\langle P_{T_1}s_{T_2} \rangle}{\langle P_{T_1}s_{T_1}
\rangle}=\frac{\overline{\langle P_{T_1}s_{T_2}
\rangle}}{\overline{\langle P_{T_1}s_{T_1} \rangle}}=\frac{\langle
P_{T_2}s_{T_2} \rangle}{\langle P_{T_2}s_{T_1}
\rangle}=\frac{\overline{\langle P_{T_2}s_{T_2}
\rangle}}{\overline{\langle P_{T_2}s_{T_1} \rangle}}= -\frac{{\rm
Im}[A_L^*A_R]}{{\rm Re}[A_L^*A_R]}\ . $$

Techniques for the measurement of the transverse polarization of
the positron have already been developed and employed for deriving
the Michel parameters \cite{transeversemichel}. Measuring the
linear polarization of the photon is going to be more challenging
but is in principle possible \cite{nim}.

In the following, we discuss the LFV process $\mu^+\to e^+e^-e^+$.
The effective Lagrangian shown in Eq. \ref{LFV-lagrangian} can also
give rise to LFV rare decay $\mu^+ \to e^+ e^-e^+$ through penguin
diagrams. Moreover, the process can also receive contributions from
the LFV four-fermion terms of the form $$ C_i \bar{\mu} \Gamma_i^\mu
(a_i P_L +b_i P_R)e \bar{e} \Gamma_{i,\mu} (c_i P_L +d_i P_R)e$$
where  $a_i$, $b_i$, $c_i$ and $d_i$ are numbers of order one and
$\Gamma_{i,\mu}=\gamma_\mu$ or 1. In the framework of R-parity
conserving MSSM which is the focus of the present study, the
couplings of the four-fermion interaction are suppressed; {\it
i.e.,} $m_\mu^2
 C_i  \ll
A_{L,R}$. Moreover, the contributions of the $A_L$ and $A_R$ terms
for the case that the momentum of one of the positrons is close to
$m_\mu/2$  is dramatically enhanced because in this limit the
virtual photon in the corresponding diagram goes on-shell. In
\cite{asli}, it is shown that by studying the transverse
polarization of the positron whose energy is close to $m_\mu/2$,
one can extract information on the phases of the underlying
theory. The maximum energy of the positrons emitted in the decay
$\mu^+ \to e^+e^-e^+$ is $E_{max}\simeq m_\mu/2-3m_e^2/(2m_\mu)$.
Consider the case that one of the positrons, $e^+_1$, has an
energy close to $E_{max}$; {\it i.e.,} $E_{max}-\Delta
E<E_1<E_{max}$ where $\Delta E \ll m_\mu$. Following \cite{asli},
let us define \be \label{Gamma-maximum} \frac{d\Gamma^{Max}}{d\cos
\theta d \phi}=\sum_{s_{e^+_2},s_{e^-} } \int_{E_{max}-\Delta
E}^{E_{max}}\int \frac{d\Gamma(\mu^+\to e^+_1e^-e^+_2)}{dE_2dE_1
d\cos\theta d\phi} dE_2 dE_1\ ,\ee where $\theta$ is the angle
between  the spin of the muon and the momentum of $e^+_1$ (the
positron whose energy is close to $E_{max}$) and $\phi$ is the
azimuthal angle that the momentum of $e^+_2$ makes with the plane
made by the momentum of $e^+_1$ and the spin of the muon.
%The sum, as indicated in the formula, is on
%the spins of $e^-$ and $e^+_2$ but not on that of $e^+_1$. Let us
%define
%$$\Gamma^{Max}\equiv \sum_{s_{e^+_1}}\int\int (d \Gamma^{Max})/(d
%\cos \theta d\phi) d \cos \theta d \phi\ . $$ $\Gamma^{Max}$ is
%the rate of $ \mu^+ \to e_1^+e^-e^+_2$ with $E_{max}-\Delta
%E<E_1<E_{max}$.
Let us suppose that a cut is employed that picks
up only events with $E_1$ within $(E_{max},E_{max}-\Delta E)$
where $2 m_e<\Delta E\ll m_\mu$. Because of the enhancement of the
amplitude at $E_1 \to E_{max}$,  the number of events passing the
cut is still significant: {\it i.e.,}
$\Gamma^{Max}/\Gamma_{tot}(\mu^+\to e^+e^-e^+)=\log (m_\mu \Delta
E/4 m_e^2)/\left( \log (m_\mu^2/4 m_e^2)-7/12 \right)>50 \% . $ As
shown in \cite{asli},
 \be \label{gammamax}\frac{d
\Gamma^{Max}}{d \cos \theta d \phi}=\frac{\alpha m_\mu}{192 \pi^3}
\left[ |A_L|^2|c_e|^2(1+\mathbb{P}_\mu \cos
\theta)+|A_R|^2|d_e|^2(1-\mathbb{P}_\mu \cos \theta)\right.\ee
$$\left. + \mathbb{P}_\mu \sin \theta \left( \cos (2 \phi) {\rm Re}[A_R
A_L^* d_e c_e^*] +\sin (2 \phi) {\rm Im}[A_R A_L^* d_e
c_e^*]\right) \right]\log \frac{m_\mu \Delta E}{4 m_e^2}, $$ where
$\mathbb{P}_\mu$ is the polarization of the initial muon  and
$c_e$ and $d_e$ are the elements of the spinor of $e^+_1$:
$v_{e^+_1}=\sqrt{2E_1} (0,d_e,c_e,0)^T$ where
$(|d_e|^2+|c_e|^2)^{1/2}=1$ and the $z$-direction is taken to be
along the momentum of $e_1^+$. Using the above formula it is
straightforward to show that the transverse polarization of
$e^+_1$ is
$$\langle s_{T_1}\rangle={\mathbb{P}_\mu \sin \theta \left( \cos 2
\phi {\rm Re}[A_R A_L^*] +\sin 2 \phi {\rm Im}[A_R A_L^*]\right)
\over  |A_L|^2(1+\mathbb{P}_\mu \cos
\theta)+|A_R|^2(1-\mathbb{P}_\mu \cos \theta)}\  $$ and
$$\langle s_{T_2}\rangle={\mathbb{P}_\mu \sin \theta \left( -\cos 2
\phi {\rm Im}[A_R A_L^*] +\sin 2 \phi {\rm Re}[A_R A_L^*]\right)
\over  |A_L|^2(1+\mathbb{P}_\mu \cos
\theta)+|A_R|^2(1-\mathbb{P}_\mu \cos \theta)}\ , $$ where
$\hat{T}_2=-(\vec{s}_\mu \times \vec{p}_{e^+_1})/|\vec{s}_\mu
\times \vec{p}_{e^+_1}|$ and $\hat{T}_1=(\hat{T}_2\times
\vec{p}_{e^+_1})/|\hat{T}_2\times \vec{p}_{e^+_1}|$.

% Let us define
%$$\tilde{S}_{T_i}(\phi_1,\phi_2)={\int_{-1}^{+1}\int_{\phi_1}^{\phi_2}
%\langle S_{T_i} \rangle \left(\sum_{s_{e^+_1}} \frac{d
%\Gamma^{Max}}{d \cos \theta d\phi}\right) d\cos \theta d \phi
%\over\int_{-1}^{+1}\int_{\phi_1}^{\phi_2}\left(\sum_{s_{e^+_1}}\frac{d
%\Gamma^{Max}}{d \cos \theta d\phi} \right) d\cos \theta d \phi} \
%.$$ $\tilde{S}_{\hat{T}_i}(\phi_1,\phi_2)$ is the average of the
%transverse polarization for decays that momentum of $e^+_2$ makes
%an azimuthal angle between $\phi_1$ and $\phi_2$ relative with the
%plane of the muon spin and the momentum of $e^+_1$. In principle,
%if the statistics is enough, $\tilde{S}_{\hat{T}_i}$ can be
%measured in the labs with already existent techniques. It is
%straightforward to verify that
%$$\tilde{S}_{T_1}(0,\frac{\pi}{2})=-\tilde{S}_{T_1}(\frac{\pi}{2},\pi)=\tilde{S}_{T_1}(\pi
%,\frac{3\pi}{2})=-\tilde{S}_{T_1}(\frac{3\pi}{2},2\pi)={\mathbb{P}_\mu
%{\rm Im}[A_RA_L^*] \over 8(|A_L|^2+|A_R|^2)}$$ and
%$$\tilde{S}_{T_2}(0,\frac{\pi}{2})=-\tilde{S}_{T_2}(\frac{\pi}{2},\pi)=\tilde{S}_{T_2}(\pi
%,\frac{3\pi}{2})=-\tilde{S}_{T_2}(\frac{3\pi}{2},2\pi)={\mathbb{P}_\mu
%{\rm Re}[A_RA_L^*] \over 8(|A_L|^2+|A_R|^2)}\ . $$ By measuring
%these quantities, we can derive the relative phase of $A_L$ and $A_R$.
Notice that the averages of $\langle s_{{T}_1}\rangle$ and
$\langle s_{{T}_2}\rangle$ over $\phi$ vanish, so to extract
information on $\arg [A_RA_L^*]$, one has to measure the azimuthal
angle that the momentum of the second positron makes with the
plane made by $\vec{s}_\mu$ and $\vec{p}_{e^+_1}$. However,
measuring $\phi$ will be challenging because when
$(P_\mu-P_{e^+_1})^2\to 0$, the angle between the momenta of the
two emitted positrons converges to $\pi$. For general
configuration with $(P_\mu-P_{e^+_{1,2}})^2\sim m_\mu^2$, the
transverse polarization of the electron also carries information
on the CP-violating phases of the underlying theory. In the
framework we are studying (R-parity conserved MSSM), the rate of
$\mu^+\to e^+e^-e^+$ is small compared to the rate of $\mu^+\to
e^+ \gamma$: ${{\rm Br}(\mu^+ \to e^+ e^+e^-)/ {\rm Br}(\mu^+\to
e^+\gamma)}\simeq \frac{\alpha}{3\pi}
[\log({m_\mu^2}/{m_e^2})-{11}/{4}]\simeq 0.0061.$ Thus, even if
the present bound on $\mu \to e\gamma$ is saturated, the
statistics of $\mu \to e ee$ will be too low to perform such
measurements in the foreseeable future. For this reason, in this
paper we will not elaborate on $\mu\to eee$ any further.

%%%%%%%%%%%%%%%%%%%%%%%%%%%%%%%%%%%%%%%%%%%%%%
%%%%%%%%%%%%%%%%%%%%%%%%%%%%%%%%%%%%%%%%%%%%%%
\section{$\mu-e$ conversion \label{conversion}}
%%%%%%%%%%%%%%%%%%%%%%%%%%%%%%%%%%%%%%%%%%%%%%
%%%%%%%%%%%%%%%%%%%%%%%%%%%%%%%%%%%%%%%%%%%%%%

 In the range of parameter space that we
are interested in, the dominant contribution to the $\mu-e$
conversion comes from the $\gamma$ and $Z$ boson exchange penguin
diagrams  and the effects of four-Fermi LFV terms can be
neglected. The effective LFV vertex in the penguin diagrams can be
parameterized as follows \ba \label{penguin}{\cal
L}_{eff}&=&\frac{e}{\sin \theta_W\cos \theta_W m_Z^2} \sum_{q\in
\{u,d\}}(H_L \bar{e}_L\gamma^\mu\mu_L+H_R \bar{e}_R \gamma^\mu
\mu_R)(Z_L^q\bar{q}_L \gamma_\mu q_L +Z_R^q\bar{q}_R \gamma_\mu
q_R)\cr &-&\sum_{q \in \{ u,d \}} \frac{Q_q
e}{p^2}\left(B^*_L\bar{e}_L\gamma_\mu\mu_L +B^*_R \bar{e}_R
\gamma_\mu \mu_R+i\frac{A_R^*}{m_\mu} \bar{e}_L \sigma^{\mu
\nu}p_\nu \mu_R +i\frac{A_L^*}{m_\mu} \bar{e}_R \sigma^{\mu
\nu}p_\nu \mu_L \right)(\bar{q} \gamma_\mu q)+{\rm H.c.}\ea where
$p=p_\mu-p_e$ is the four-momentum transferred by the photon or
$Z$-boson and $Q_q$ is the electric charge of the quark.
$Z_{L(R)}^q=T^3_q-Q_q \sin^2 \theta_W$ is the coupling of
left(right)-handed quark to the $Z$-boson. $H_L$ and $H_R$ are the
effective couplings of the $Z$ boson to lepton. $A_L$ and $A_R$
are the same couplings that appear in Eq.~(\ref{LFV-lagrangian}).
$B_L(p^2)$ and $ B_R(p^2)$ vanish for $p^2\to 0$ so they do not
contribute to $\mu \to e \gamma$. Let us evaluate and compare the
contributions of the various couplings appearing in
Eq.~(\ref{penguin}). Since $A_L$ and $A_R$ flip the chirality,
they are suppressed by a factor of $m_\mu$. Ward identity implies
that $B_R$ and $B_L$ are suppressed by $p^2=-m_\mu^2$. There is
not such a suppression in $H_L$ and $H_R$, thus
$H_{L(R)}/m_Z^2\sim B_{L(R)}/m_\mu^2$.

\be \frac{d\Gamma(\mu N \to e N)}{d\cos \theta}=S\left[
\frac{1-\mathbb{P}_\mu \cos \theta}{2}|aH_L+b(
A_R^*+B_L^*)|^2+\frac{1+\mathbb{P}_\mu \cos \theta}{2}|aH_R+b(
A_L^*+B_R^*)|^2 \right] \ , \label{partial-conversion}\ee where
$S$ is a numerical factor that includes the nuclear form factor
\cite{hisano} and \be \label{aANDb}a= \frac{e\left[Z(1/2-2\sin^2
\theta_W)-N/2\right]}{2m_Z^2 \sin \theta_W \cos \theta_W} \ \ \ \
{\rm and} \ \ \ \ b=\frac{eZ}{m_\mu^2}\ee in which $Z$ and $N$ are
respectively the numbers of protons and neutrons inside the
nucleus.

Let us define \be \label{KR} K_R\equiv aH_L+b(A_R^*+B_L^*)\ee and
\be \label{KL} K_L\equiv aH_R+b(A_L^*+B_R^*) .\ee
 From Eq.~(\ref{partial-conversion}), we observe that
the total conversion rate, $\int (d\Gamma/d \cos\theta) d\cos
\theta$ provides us with information on the sum of $|K_R|^2$ and
$|K_L|^2$. That is while by studying the angular distribution of the
final electron, we can also extract \be R_2\equiv {|K_R|^2-|K_L|^2
\over |K_R|^2+|K_L|^2}. \label{R2}\ee
   Let us now
study what extra information can be extracted by measuring the
spin of the final electron.

Similarly to the case of $\mu \to e \gamma$, let us define the
directions $\hat{T}_1$ and $\hat{T}_2$ as follows:
$\hat{T}_2=({\vec{p}_e\times \vec{s}_\mu})/{|\vec{p}_e\times
\vec{s}_\mu|} \ \ \ {\rm and} \ \ \ \hat{T}_1=({(\vec{p}_e\times
\vec{s}_\mu)\times \vec{p}_e})/{|(\vec{p}_e\times
\vec{s}_\mu)\times \vec{p}_e|}\ .$ Let us also define
$$\langle s_{T_i} \rangle \equiv {d\Gamma\left[\mu N\to
e(\vec{s}_e=\frac{1}{2}\hat{T}_i) N\right]-d\Gamma\left[\mu N\to
e(\vec{s}_e=-\frac{1}{2}\hat{T}_i) N\right] \over \sum_{\vec{s}_e}
d\Gamma[\mu N \to e N]} \ . $$
 It is straightforward to verify
that  the transverse polarization of the emitted electron in the
directions of $\hat{T}_1$ and $\hat{T}_2$ are \be
\label{conversion-sT1}\langle s_{{T_1}} \rangle=\frac{2{\rm
Re}\left[K_RK_L^* \right]\mathbb{P}_\mu \sin \theta}{
|K_R|^2(1-\mathbb{P}_\mu \cos \theta)+|K_L|^2(1+\mathbb{P}_\mu
\cos \theta)}\ ,\ee \be \label{conversion-sT2}\langle s_{{T_2}}
\rangle=\frac{2{\rm Im}\left[K_RK_L^* \right]\mathbb{P}_\mu \sin
\theta}{ |K_R|^2(1-\mathbb{P}_\mu \cos
\theta)+|K_L|^2(1+\mathbb{P}_\mu \cos \theta)}\ .\ee  Averaging
over the angular distribution, we find \be
\label{mean-conversion-sT1}\overline{\langle s_{{T_1}}
\rangle}\equiv {\int \langle s_{T_1}\rangle \frac{d\Gamma}{d \cos
\theta} d\cos \theta \over\int\frac{d\Gamma}{d \cos \theta} d\cos
\theta}= \frac{\pi{\rm Re}\left[K_RK_L^*\right]\mathbb{P}_\mu
}{2\left(|K_R|^2+|K_L)|^2\right)}\ ,\ee and \be
\label{mean-conversion-sT2} \overline{\langle s_{{T_2}}
\rangle}\equiv {\int \langle s_{T_2}\rangle \frac{d\Gamma}{d \cos
\theta} d\cos \theta \over\int\frac{d\Gamma}{d \cos \theta} d\cos
\theta}=\frac{\pi {\rm Im}\left[K_RK_L^* \right]\mathbb{P}_\mu
}{2\left(|K_L|^2+|K_R|^2\right)}\ .\ee The advantage of the study
of the $\mu - e $ conversion over the study of $\mu \to e\gamma$
is that in the former case there is no need for performing the
challenging photon polarization measurement. The drawback is the
polarization of the initial muon. While the polarization of muon
in the $\mu \to e \gamma$ experiments is close to 100\%, the muons
orbiting the nuclei (the muons  in the $\mu-e$ conversion
experiments) suffer from low polarization of 16\% or lower
\cite{16-or-less}. However, there are proposals to ``re-"polarize
the muon in the muonic atoms by using polarized nuclear targets
\cite{repolarize}.

In this paper, we take $\mathbb{P}_\mu=20\%$. For any given value
of $\mathbb{P}_\mu$, our results can be simply re-scaled.

%%%%%%%%%%%%%%%%%%%%%%%%%%%%%%%%%%
%%%%%%%%%%%%%%%%%%%%%%%%%%%%%%%%%5
\section{Effects of the CP-violating phases of MSSM \label{impact}}
%%%%%%%%%%%%%%%%%%%%%%%%%%%%%%%%%%%%%%%5
%%%%%%%%%%%%%%%%%%%%%%%%%%%%%%%%%%%%%%%%%55

In this section, we study the polarizations introduced in the
previous section in the framework of R-parity conserving Minimal
Supersymmetric Standard Model (MSSM). The part of the superpotential
that is relevant to this study can be written as
 \be \label{superpotential}
   W_{MSSM} =
           - Y_{i}  \widehat{{e}^c_{Ri}} \ \widehat{L}_i  \cdot \widehat{H_{d}}
-\mu\ \widehat{H_{u}}\cdot \widehat{H_{d}} \ee
 where $\widehat{L}_i$, $\widehat{H_{u}}$ and $ \widehat{H_{d}}$
 are doublets of chiral superfields associated respectively with
 the left-handed lepton doublets and the two Higgs doublets of the
 MSSM.  $\widehat{{e}^c_{Ri}}$ is
the chiral superfield associated with the  right-handed charged
lepton field, $e^c_{Ri}$. The index ``$i$" is the flavor index. At
the electroweak scale, the soft supersymmetry breaking part of
Lagrangian in general has the following form \ba
\label{MSSMsoft}\L_{\rm soft}^{\rm MSSM} &=&-\ 1/2 \left( M_1
\widetilde{B}\widetilde{B}+ M_2 \widetilde{W} \widetilde{W} +{\rm
H.c.} \right) \cr &-&((A_{i}Y_{ i}\delta_{ij}+A_{ij})
\widetilde{e_{Ri}^c} \ \widetilde{L_{j}} \cdot H_{d} + {\rm H.c.} )
- \widetilde{L_{i}} ^{\dag} \ ( m_{L }^{2})_{ij}\widetilde{L_{j}} -
\widetilde{e_{Ri}^c }^{\dag} \ (m_{R
}^{2})_{ij}\widetilde{e_{Rj}^c}\cr &-& \ m_{H_{u}}^{2}\
H_{u}^{\dag}\ H_{u}-\ m_{H_{d}}^{2}\ H_{d}^{\dag}\ H_{d}-(\ B_H \
H_{u}\cdot H_{d}+ {\rm H.c.}),\ea where the ``$i$" and ``$j$"
indices determine the flavor and $\tilde{L}_i$ consists of
$(\tilde{\nu}_i \ \tilde{e}_{Li})$. Notice that we have divided the
trilinear coupling to a diagonal flavor part
($A_{i}Y_{e_i}\delta_{ij}$) and a LFV part ($A_{ij}$ with
$A_{ii}=0$).  Terms involving the squarks as well as the gluino mass
term have to be added to Eqs.~(\ref{superpotential},\ref{MSSMsoft})
but these terms are not relevant to this study. The Hermiticity of
the Lagrangian implies that $m_{H_{u}}^2$, $m_{H_{d}}^2$, and the
diagonal elements of $m_{L}^2$ and $m_{{R }}^2$ are all real.
Moreover, without loss of generality, we can rephase the fields to
make the parameters $M_2$, $B_H$ as well as $Y_{ i}$ real. In such a
basis, the rest of the above parameters can in general be complex
and can be considered as sources of CP-violation. After electroweak
symmetry breaking, $A_{ij}$ gives rise to LFV masses:
$$(m^2_{LR})_{ij}=A_{ij}\langle H_d\rangle \ \ \ {\rm for}\ i\ne j \
.$$ Notice that in general $|A_{ij}|\ne|A_{ji}|$ and therefore
$|(m^2_{LR})_{ij}|\ne |(m^2_{LR})_{ji}|$.

\begin{figure}
\begin{center}
\centerline{\vspace{-1.2cm}}
\centerline{\includegraphics[scale=0.4]{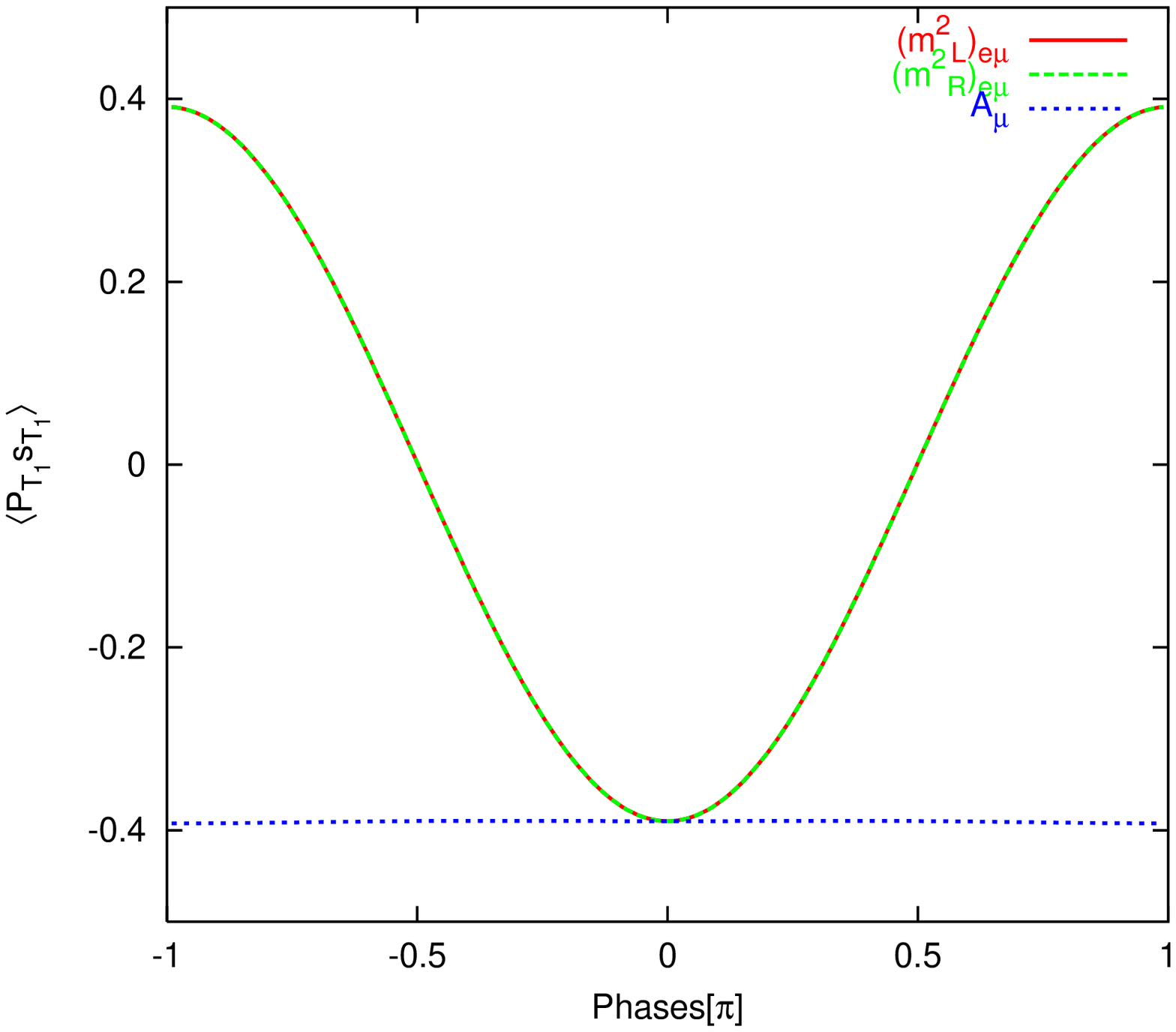}\hspace{5mm}\includegraphics[scale=0.4]{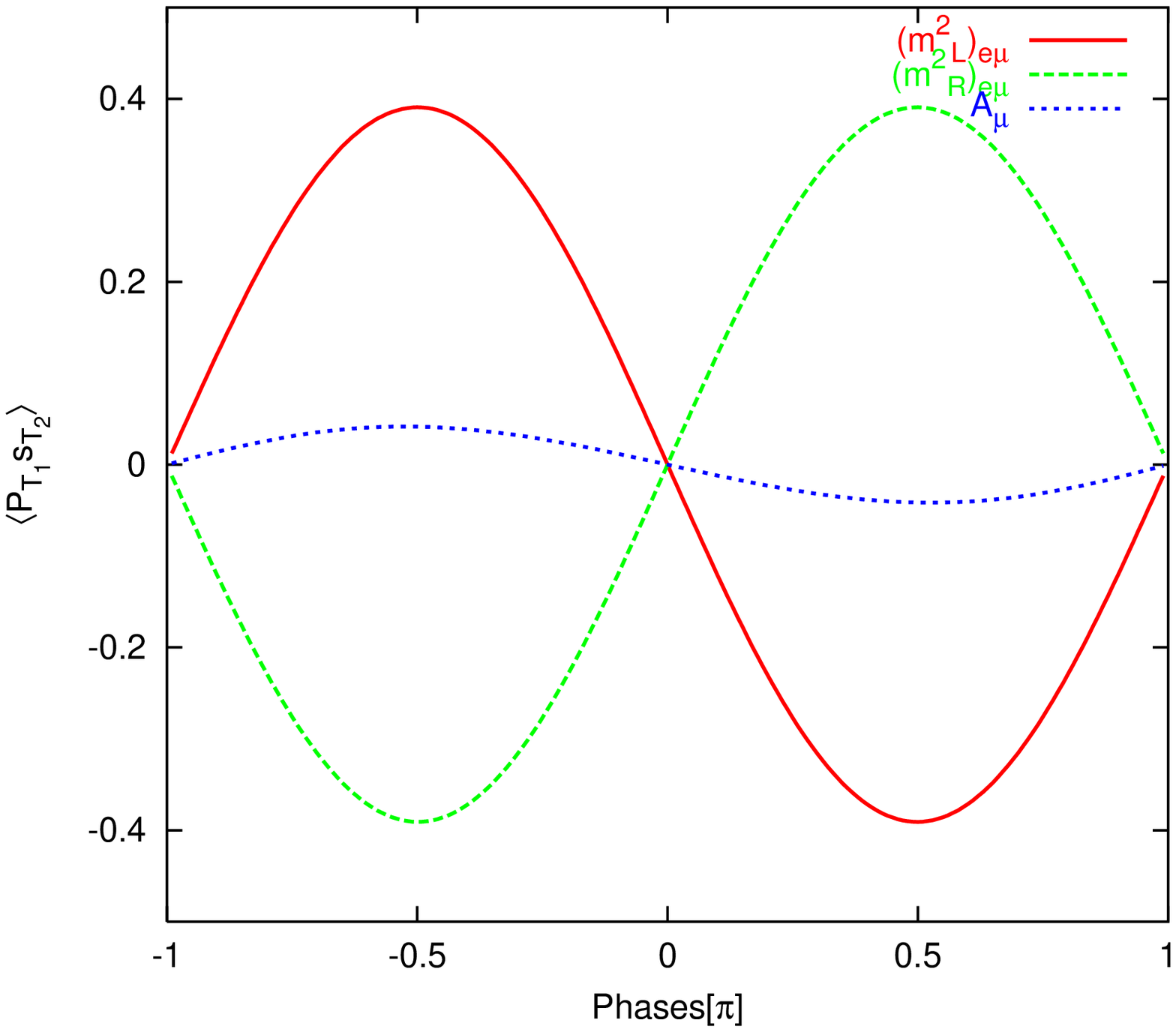}}
\centerline{\vspace{1.cm}\hspace{1cm}(a)\hspace{7cm}(b)}
\centerline{\vspace{-1.2cm}}
\centerline{\includegraphics[scale=0.4]{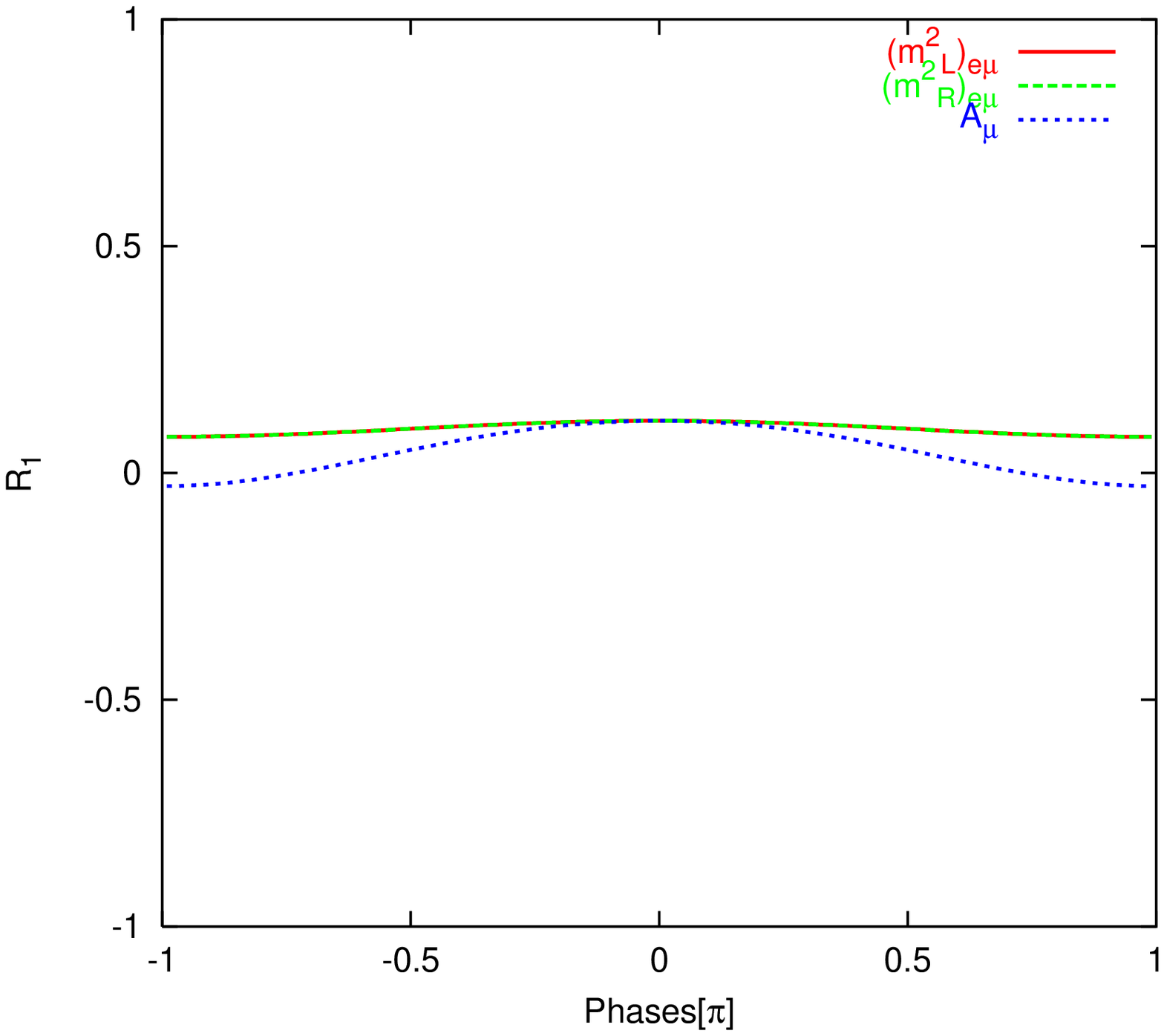}\hspace{5mm}\includegraphics[scale=0.4]{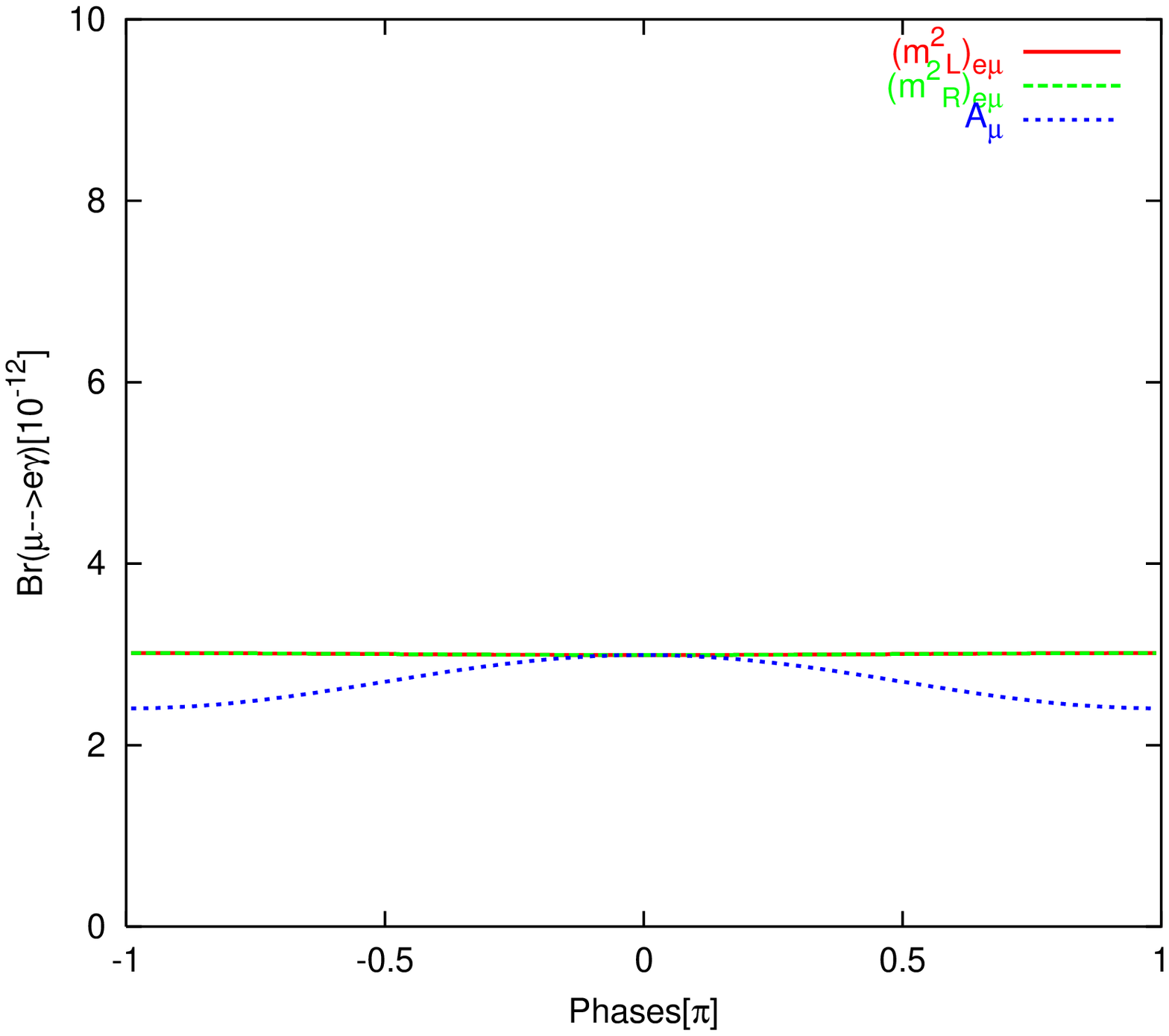}}
\centerline{\hspace{1.2cm}(c)\hspace{7cm}(d)}
\centerline{\vspace{-1.5cm}}
\end{center}
\caption{ Observable quantities in the $\mu \to e\gamma$
experiment versus the phases of $A_\mu$, $(m^2_L)_{e\mu }$ and
$(m^2_R)_{e\mu }$. The vertical axes  in Figs. (a)-(d) are
respectively $\overline{\langle P_{T_1}s_{T_1}}\rangle$,
$\overline{\langle P_{T_1}s_{T_2}}\rangle$, $R_1$  and ${\rm
Br}(\mu\rightarrow e\gamma)$. The input parameters correspond to
the $P3$ benchmark proposed in \cite{Heinemeyer:2007cn}:
$|\mu|=400$~GeV, $m_0=1000$~GeV, ${\rm M}_{1/2}=500$~GeV and $\tan
\beta=10$ and we have set $|A_{\mu}|$=$|A_{e}|$=700~GeV. All the
LFV elements of the slepton mass matrix are set to zero except
$(m_L^2)_{e \mu}=2500~{\rm GeV}^2$ and $(m_R^2)_{e \mu}=12500~{\rm
GeV}^2$. We have taken $\mathbb{P}_\mu=100 \%$.
}\label{sim-muegamma}
\end{figure}

\begin{figure}
\begin{center}
\centerline{\vspace{-1.2cm}}
\centerline{\includegraphics[scale=0.4]{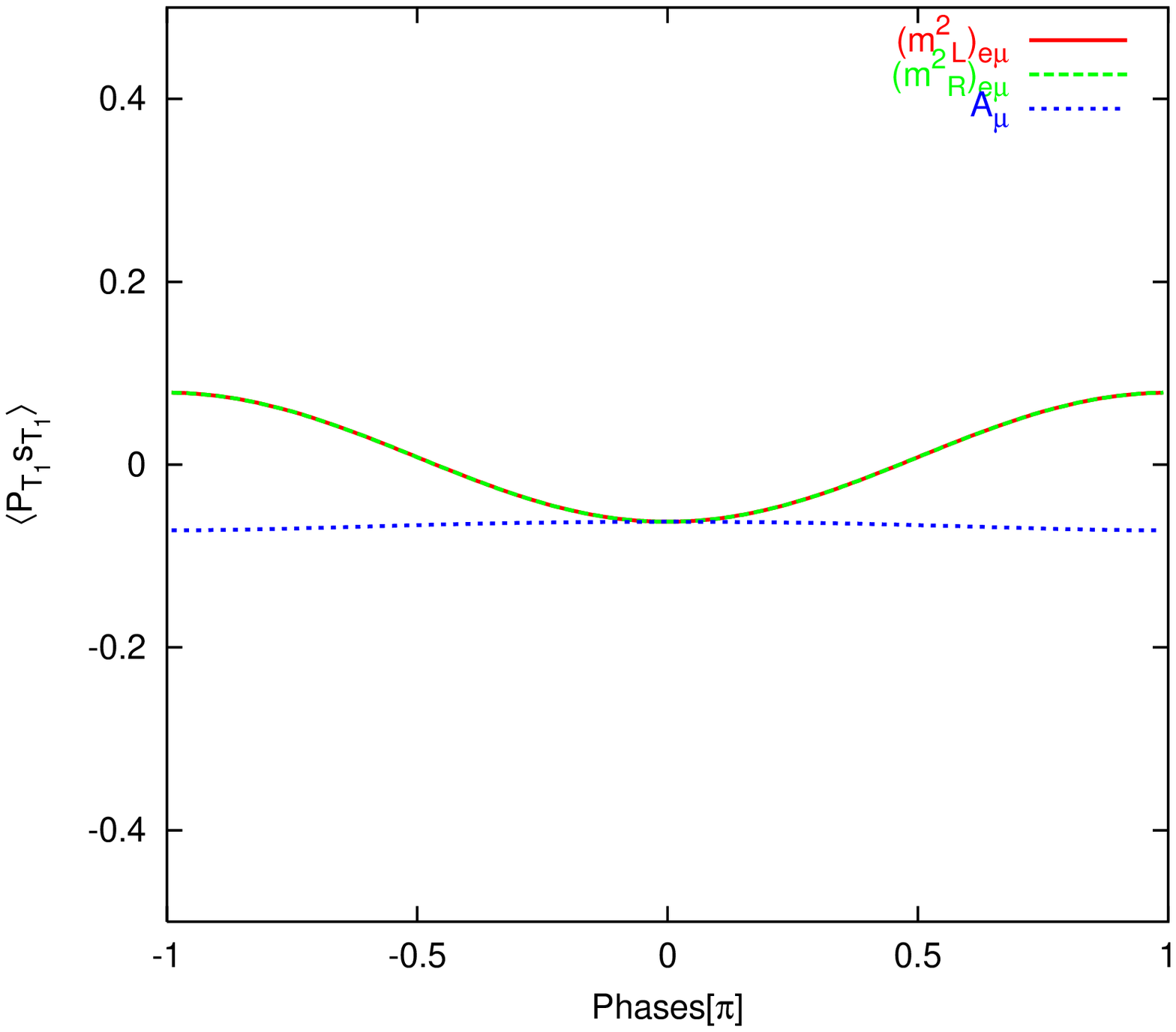}\hspace{5mm}\includegraphics[scale=0.4]{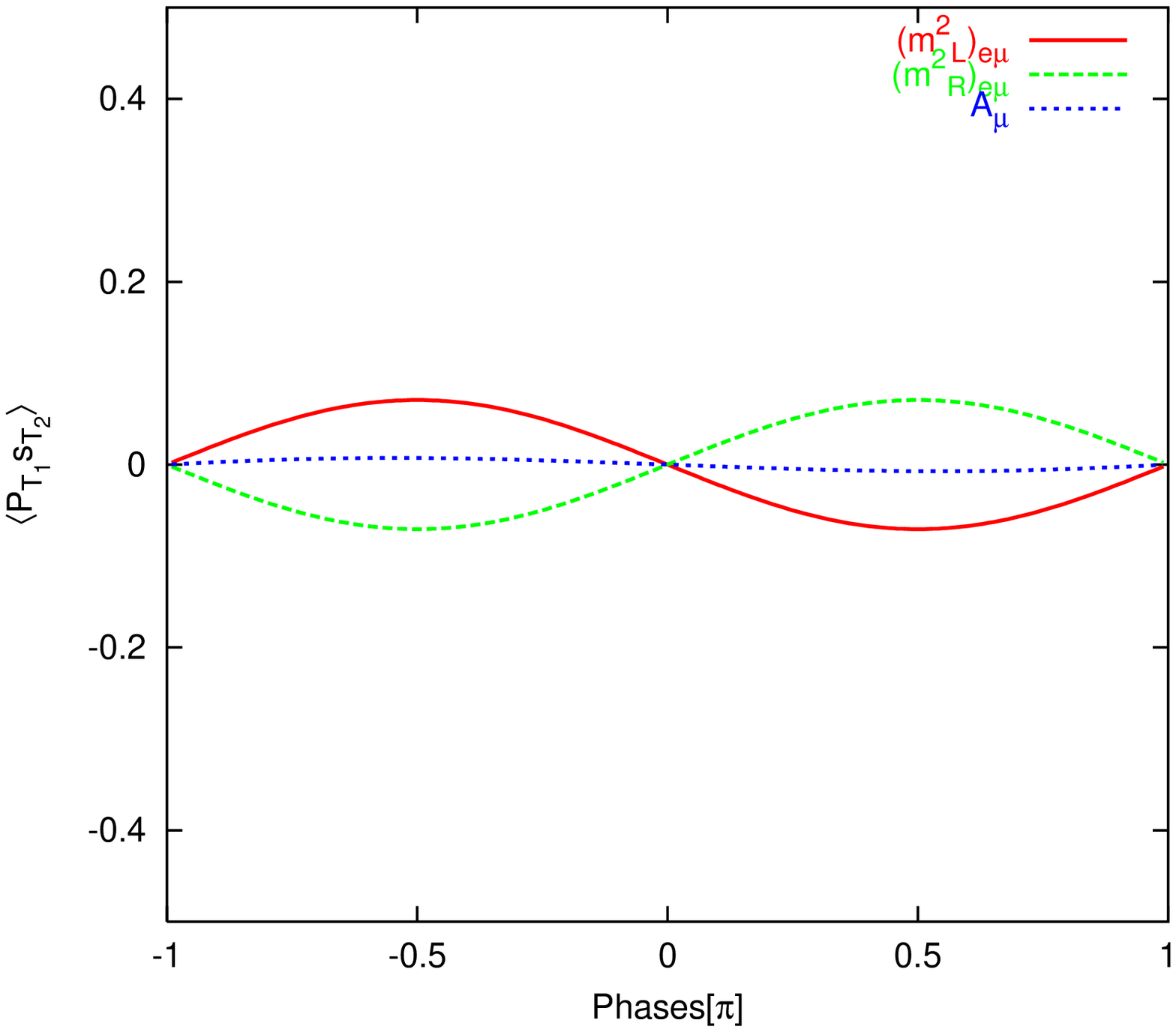}}
\centerline{\vspace{1.cm}\hspace{1cm}(a)\hspace{7cm}(b)}
\centerline{\vspace{-1.2cm}}
\centerline{\includegraphics[scale=0.4]{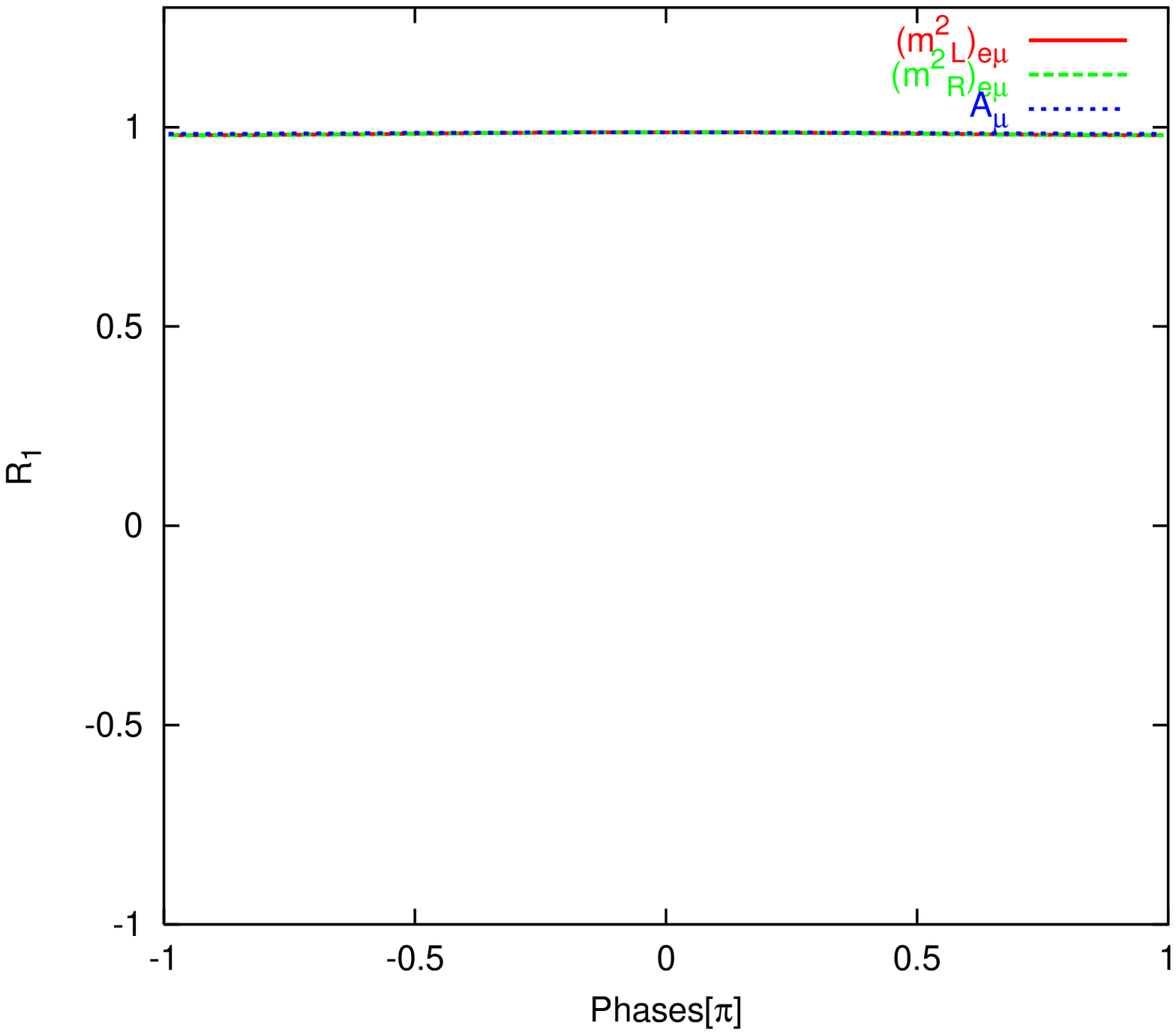}\hspace{5mm}\includegraphics[scale=0.4]{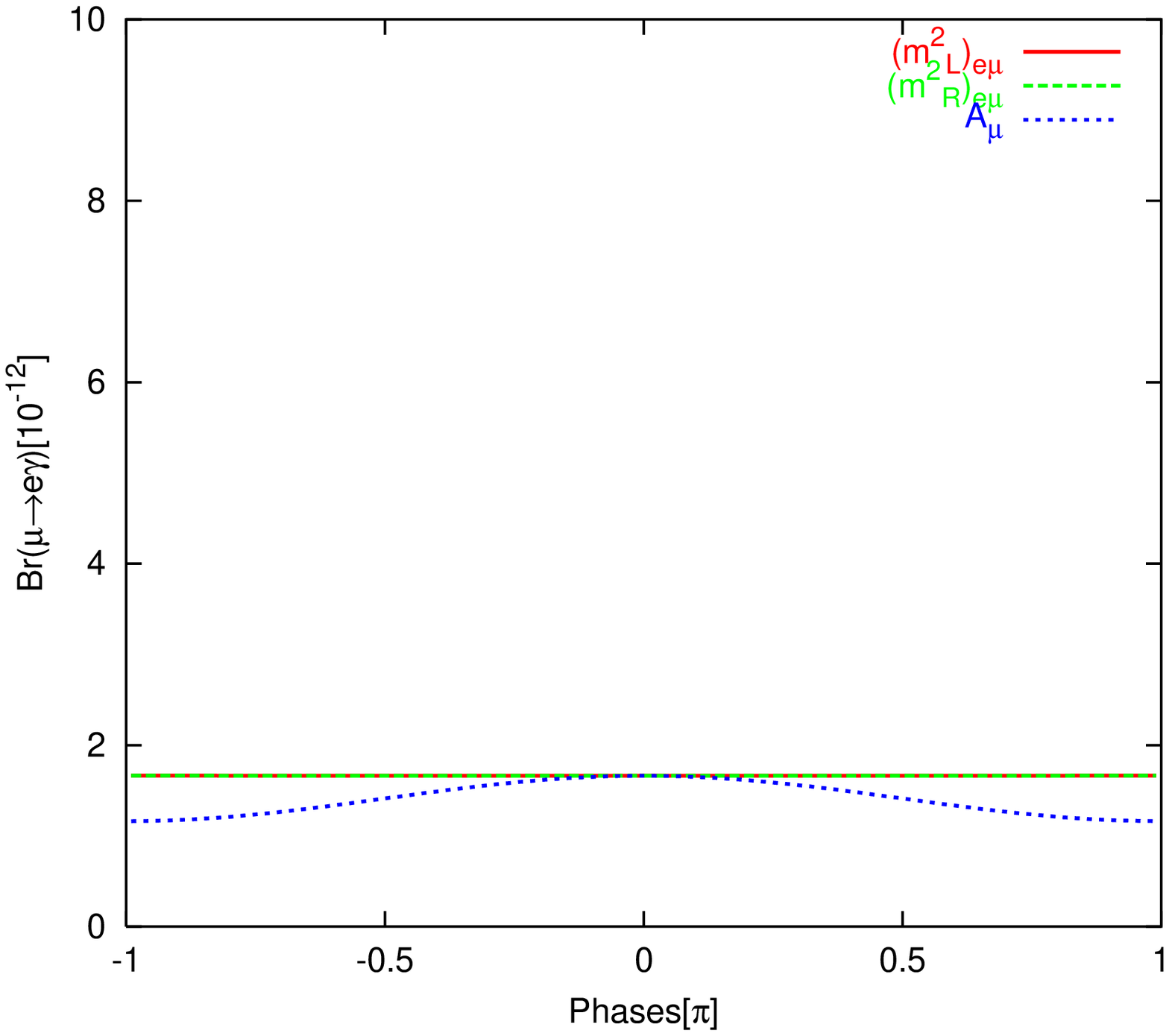}}
\centerline{\hspace{1.2cm}(c)\hspace{7cm}(d)}
\centerline{\vspace{-1.5cm}}
\end{center}
\caption{ Observable quantities in the $\mu \to e\gamma$
experiment versus the phases of $A_\mu$, $(m^2_L)_{e\mu }$ and
$(m^2_R)_{e\mu }$. The vertical axes  in Figs. (a)-(d) are
respectively $\overline{\langle P_{T_1}s_{T_1}}\rangle$,
$\overline{\langle P_{T_1}s_{T_2}}\rangle$, $R_1$  and ${\rm
Br}(\mu\rightarrow e\gamma)$.  The input parameters correspond to
the $P3$ benchmark proposed in \cite{Heinemeyer:2007cn}:
$|\mu|=400$~GeV, $m_0=1000$~GeV, ${\rm M}_{1/2}=500$~GeV and $\tan
\beta=10$ and we have set $|A_{\mu}|$=$|A_{e}|$=700~GeV. All the
LFV elements of the slepton mass matrix are set to zero except
$(m_L^2)_{e \mu}=250~{\rm GeV}^2$ and $(m_R^2)_{e \mu}=12500~{\rm
GeV}^2$. We have taken $\mathbb{P}_\mu=100
\%$.}\label{hier-muegamma}
\end{figure}

\begin{figure}
\begin{center}
\centerline{\vspace{-1.2cm}}
\centerline{\includegraphics[scale=0.4]{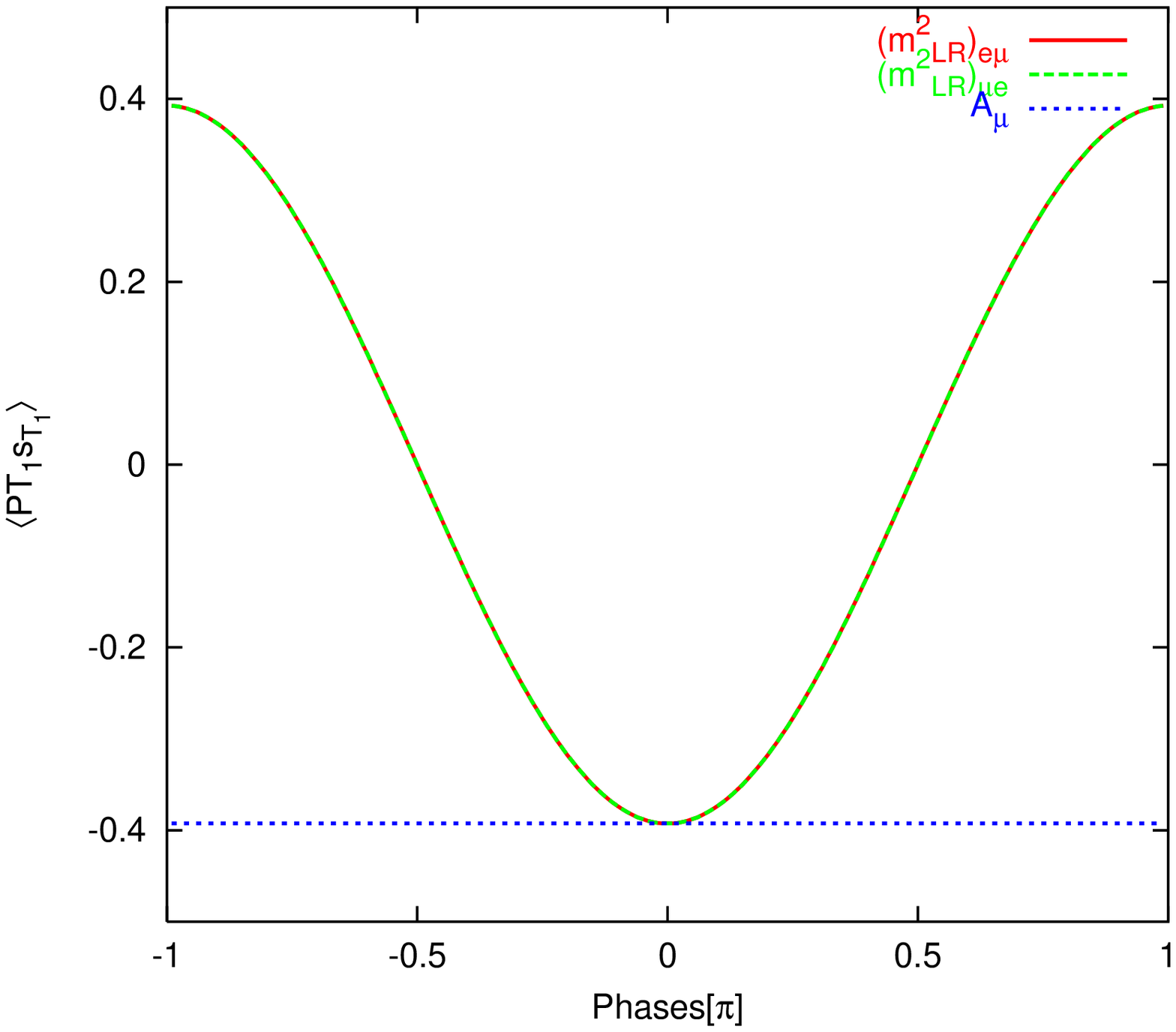}\hspace{5mm}\includegraphics[scale=0.4]{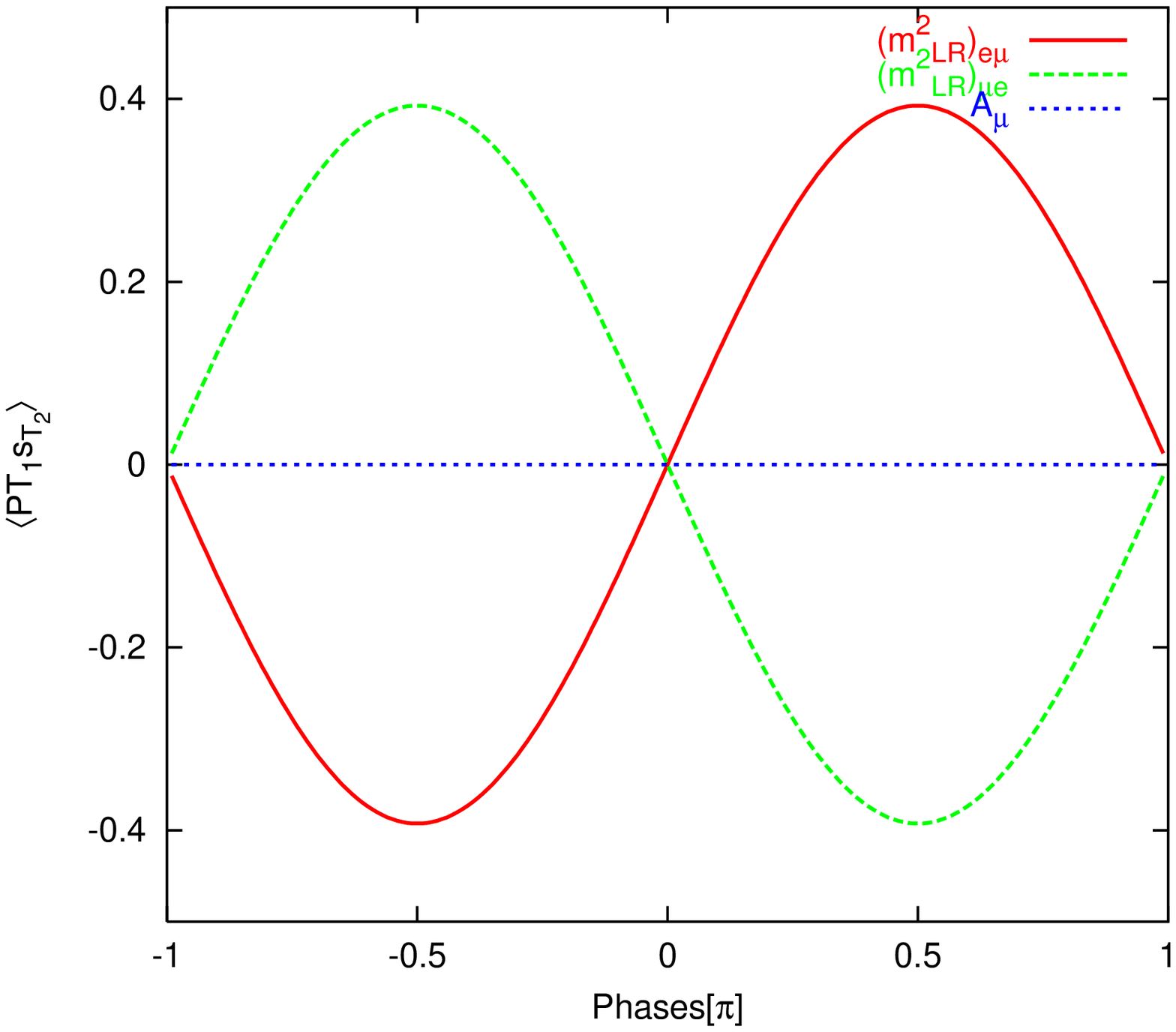}}
\centerline{\vspace{1.cm}\hspace{1cm}(a)\hspace{7cm}(b)}
\centerline{\vspace{-1.2cm}}
\centerline{\includegraphics[scale=0.4]{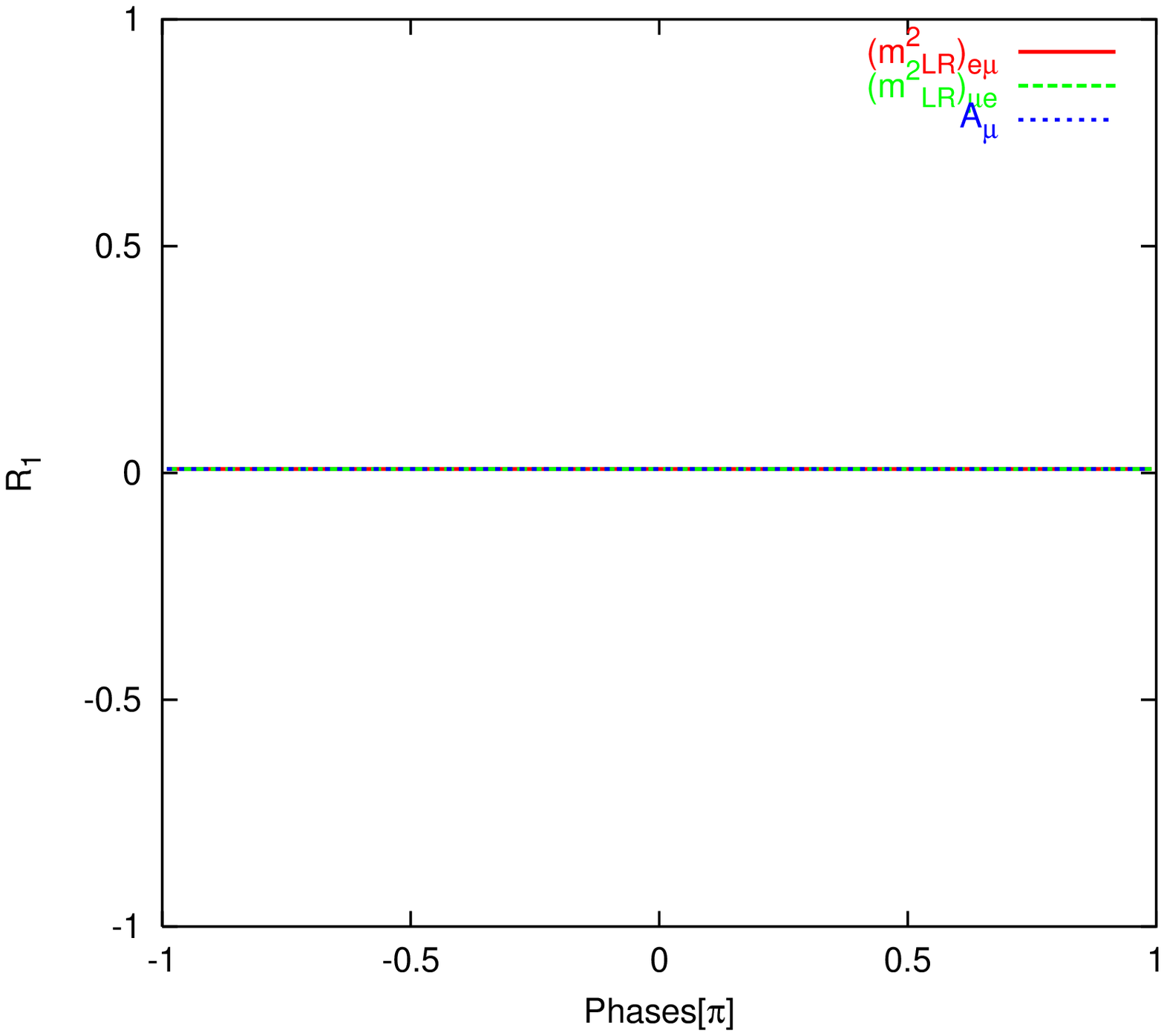}\hspace{5mm}\includegraphics[scale=0.4]{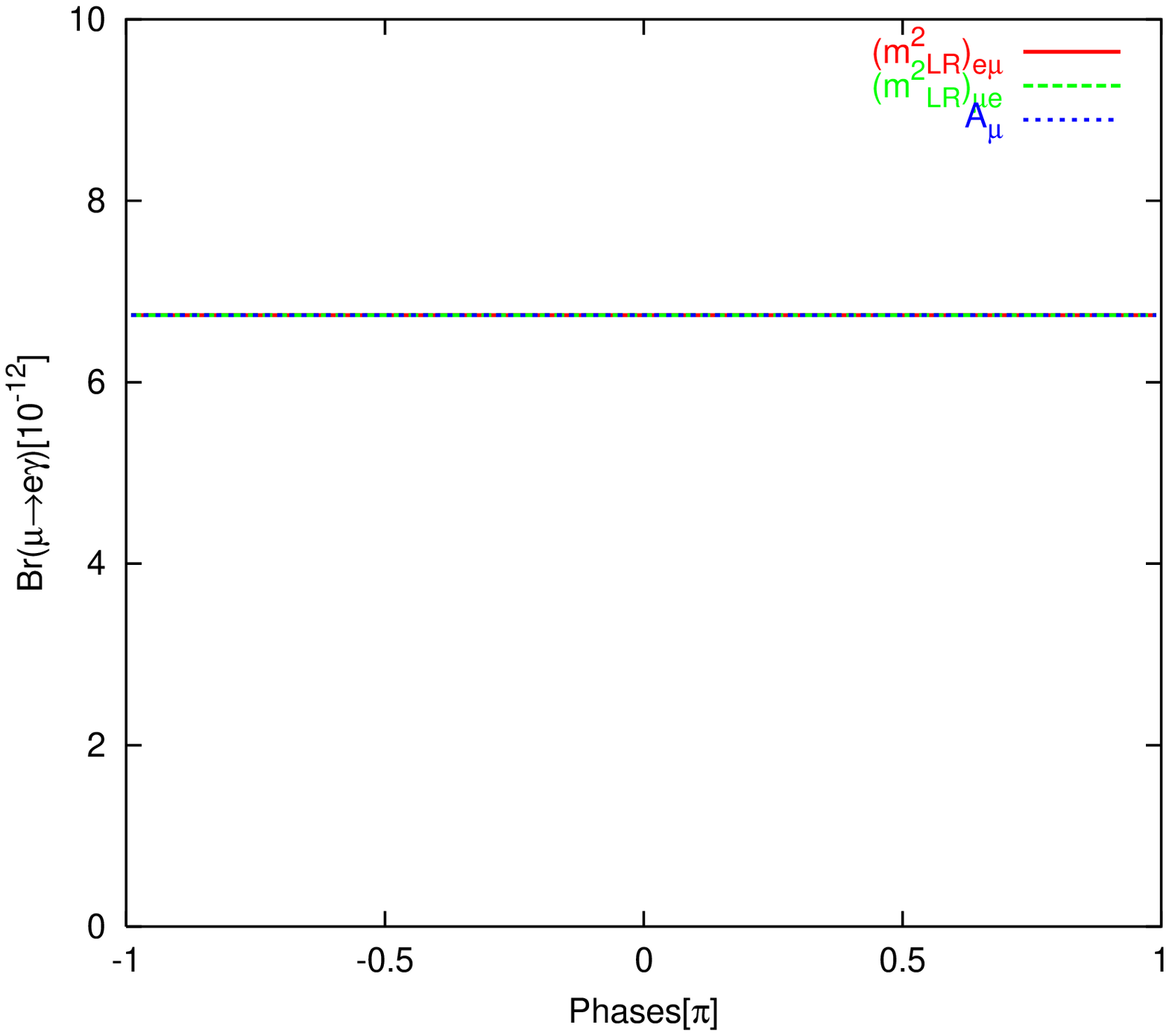}}
\centerline{\hspace{1.2cm}(c)\hspace{7cm}(d)}
\centerline{\vspace{-1.5cm}}
\end{center}
\caption{ Observable quantities in the $\mu \to e\gamma$
experiment versus the phases of $A_\mu$, $(m^2_{LR})_{e\mu }$ and
$(m^2_{LR})_{\mu e}$. The vertical axes  in Figs. (a)-(d) are
respectively $\overline{\langle P_{T_1}s_{T_1}}\rangle$,
$\overline{\langle P_{T_1}s_{T_2}}\rangle$, $R_1$  and ${\rm
Br}(\mu\rightarrow e\gamma)$.  The input parameters correspond to
the $P3$ benchmark proposed in \cite{Heinemeyer:2007cn}:
$|\mu|=400$~GeV, $m_0=1000$~GeV, ${\rm M}_{1/2}=500$~GeV and $\tan
\beta=10$ and we have set $|A_{\mu}|$=$|A_{e}|$=700~GeV. All the
LFV elements of the slepton mass matrix are set to zero except
$(m_{LR}^2)_{e\mu}(=A_{e\mu}\langle H_d\rangle)$=$14~{\rm GeV}^2$
and $(m_{LR}^2)_{\mu e}(=A_{\mu e}\langle H_d\rangle)$=$14~{\rm
GeV}^2$. We have taken $\mathbb{P}_\mu=100 \%$.}\label{LRdecay}
\end{figure}

\begin{figure}
\begin{center}
\centerline{\vspace{-1.2cm}}
\centerline{\includegraphics[scale=0.4]{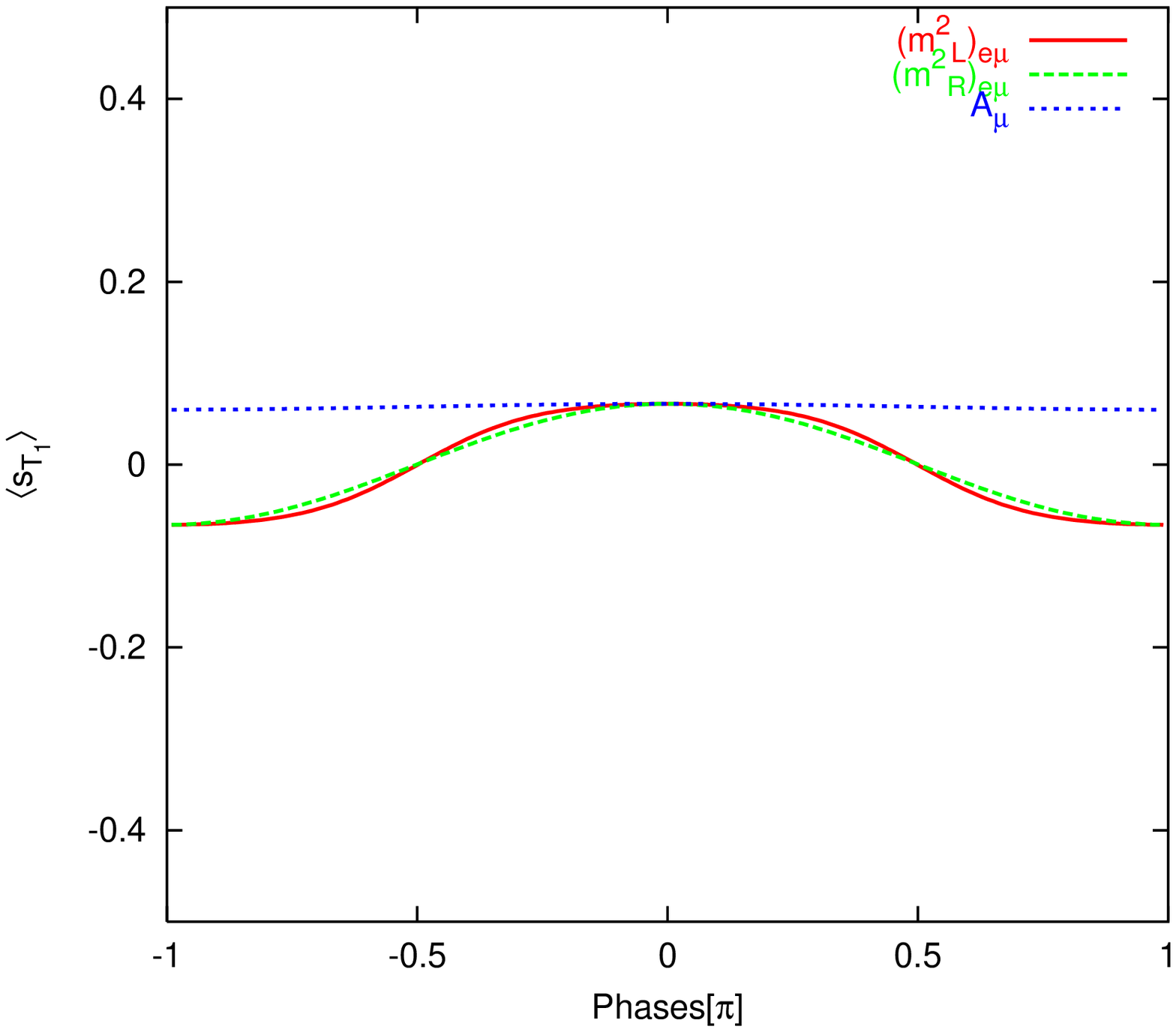}\hspace{5mm}\includegraphics[scale=0.4]{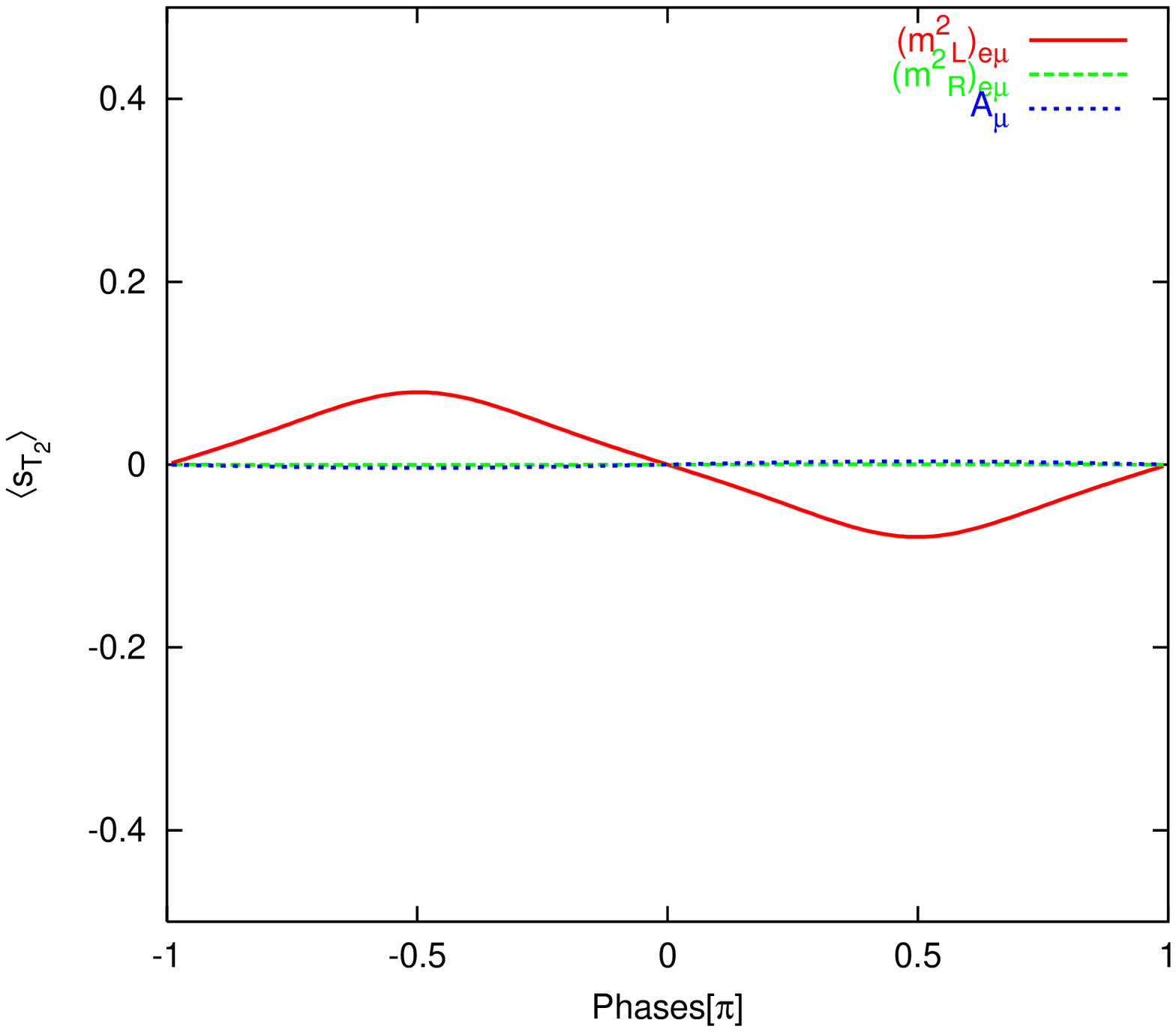}}
\centerline{\hspace{1.2cm}(a)\hspace{7cm}(b)}
\centerline{\includegraphics[scale=0.4]{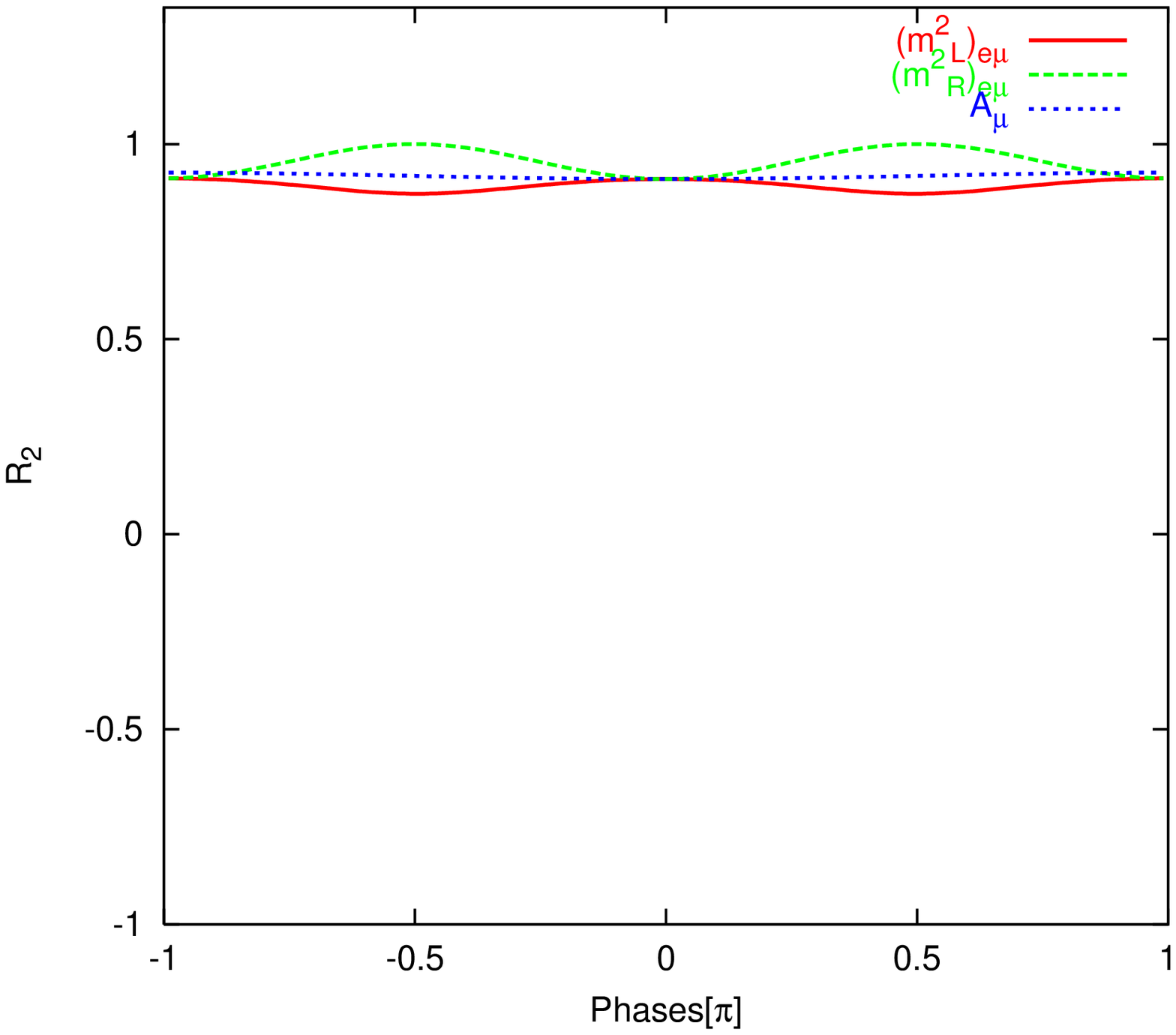}\hspace{5mm}\includegraphics[scale=0.4]{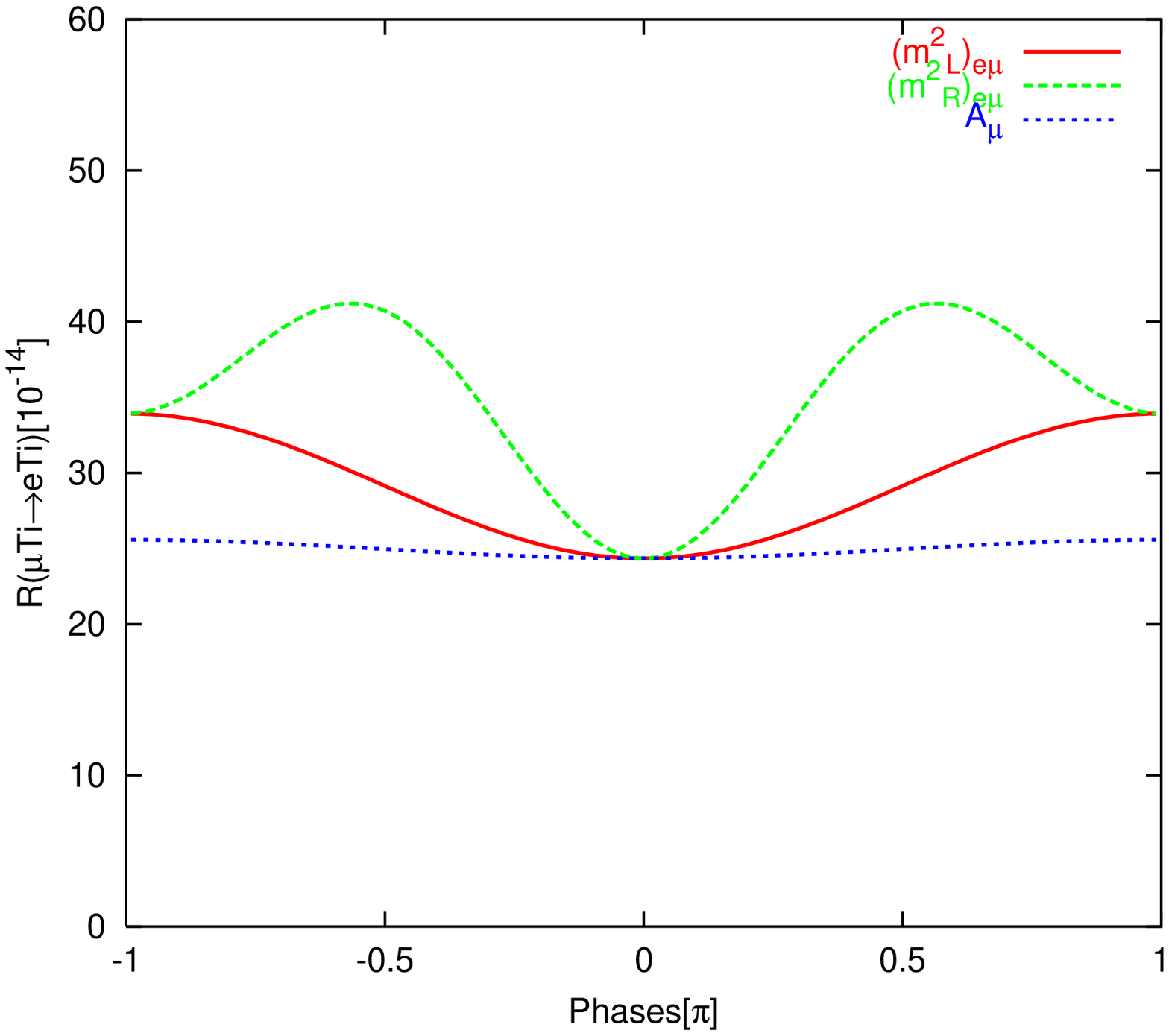}}
\centerline{\hspace{1.2cm}(c)\hspace{7cm}(d)}
\centerline{\vspace{-1.5cm}}
\end{center}
\caption{ Observable quantities in the $\mu- e$ conversion
experiment versus the phases of $A_\mu$, $(m^2_L)_{e\mu }$ and
$(m^2_R)_{e\mu }$. The vertical axes  in Figs. (a)-(d) are
respectively
  $\overline{\langle s_{T_1}}\rangle$, $\overline{\langle
s_{T_2}}\rangle$, $R_2$  and $R(\mu {\rm Ti}\rightarrow e {\rm
Ti})$.  The input parameters correspond to the $P3$ benchmark
proposed in \cite{Heinemeyer:2007cn}: $|\mu|=400$~GeV,
$m_0=1000$~GeV, ${\rm M}_{1/2}=500$~GeV and $\tan \beta=10$ and we
have set $|A_{\mu}|$=$|A_{e}|$=700~GeV. All the LFV elements of the
slepton mass matrix are set to zero except $(m_L^2)_{e
\mu}=2500~{\rm GeV}^2$ and $(m_R^2)_{e \mu}=12500~{\rm GeV}^2$.  We
have taken $\mathbb{P}_\mu=20 \%$. }\label{sim-conversion}
\end{figure}

\begin{figure}
\begin{center}
\centerline{\vspace{-1.2cm}}
\centerline{\includegraphics[scale=0.4]{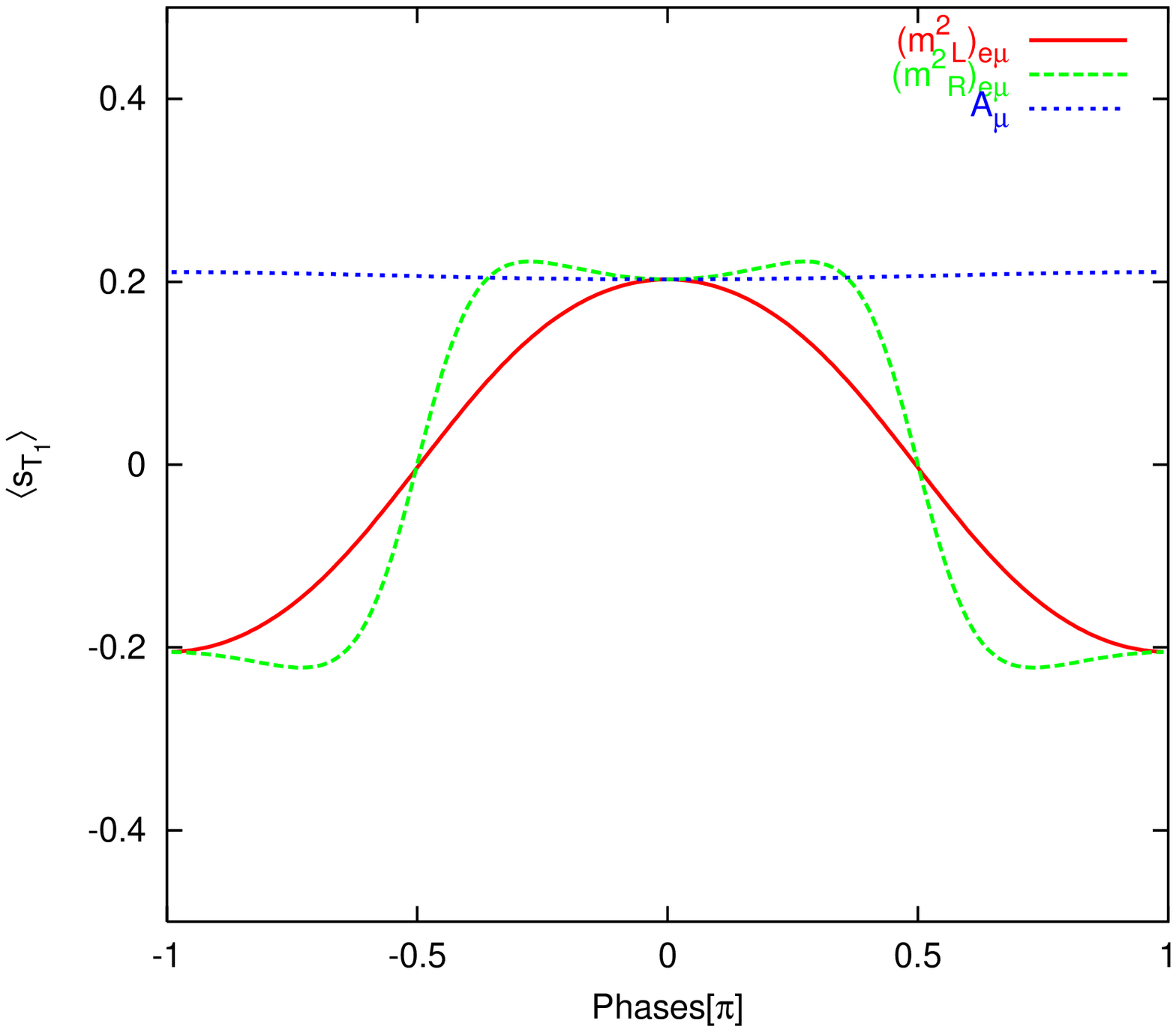}\hspace{5mm}\includegraphics[scale=0.4]{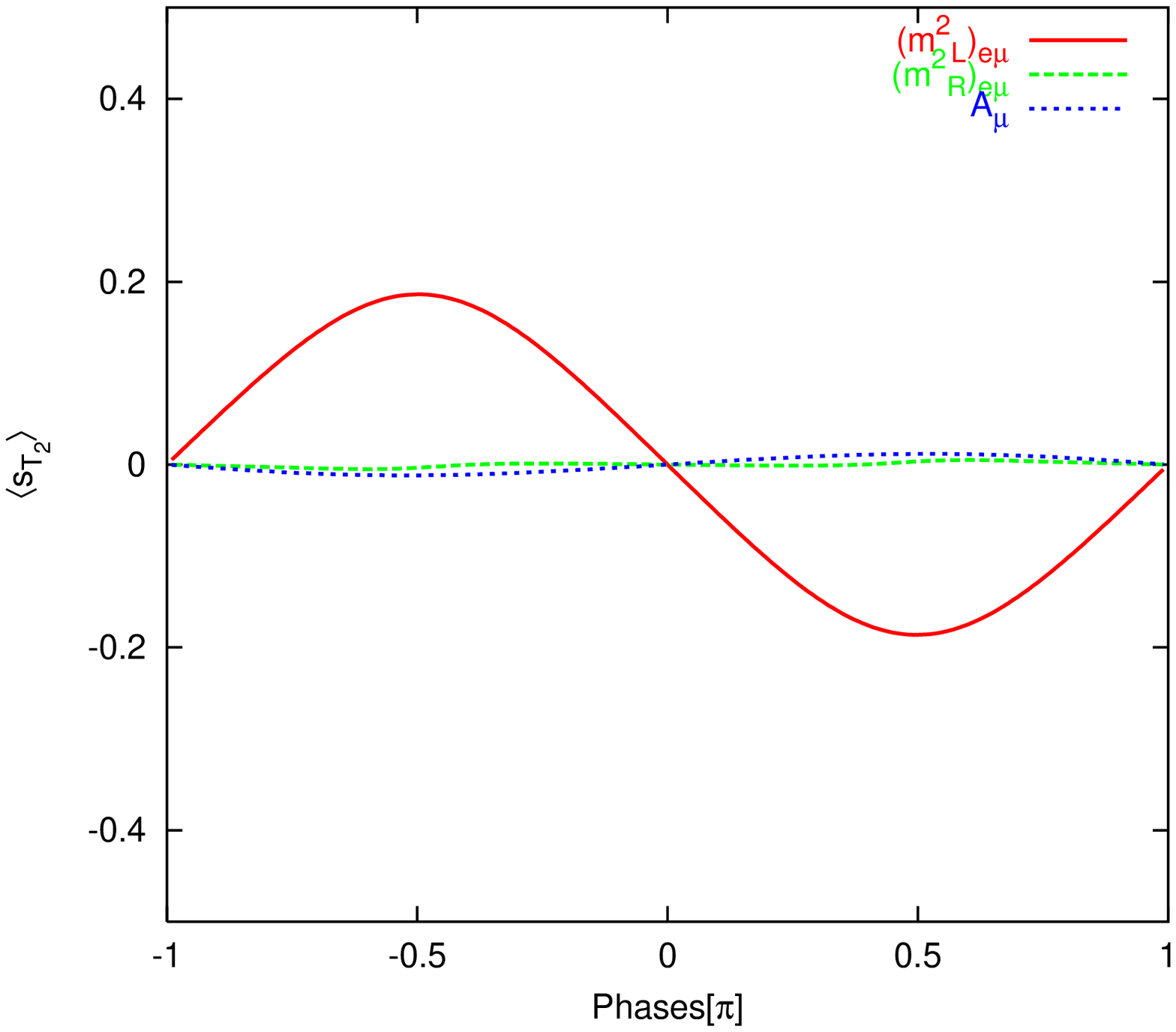}}
\centerline{\hspace{1.2cm}(a)\hspace{7cm}(b)}
\centerline{\includegraphics[scale=0.4]{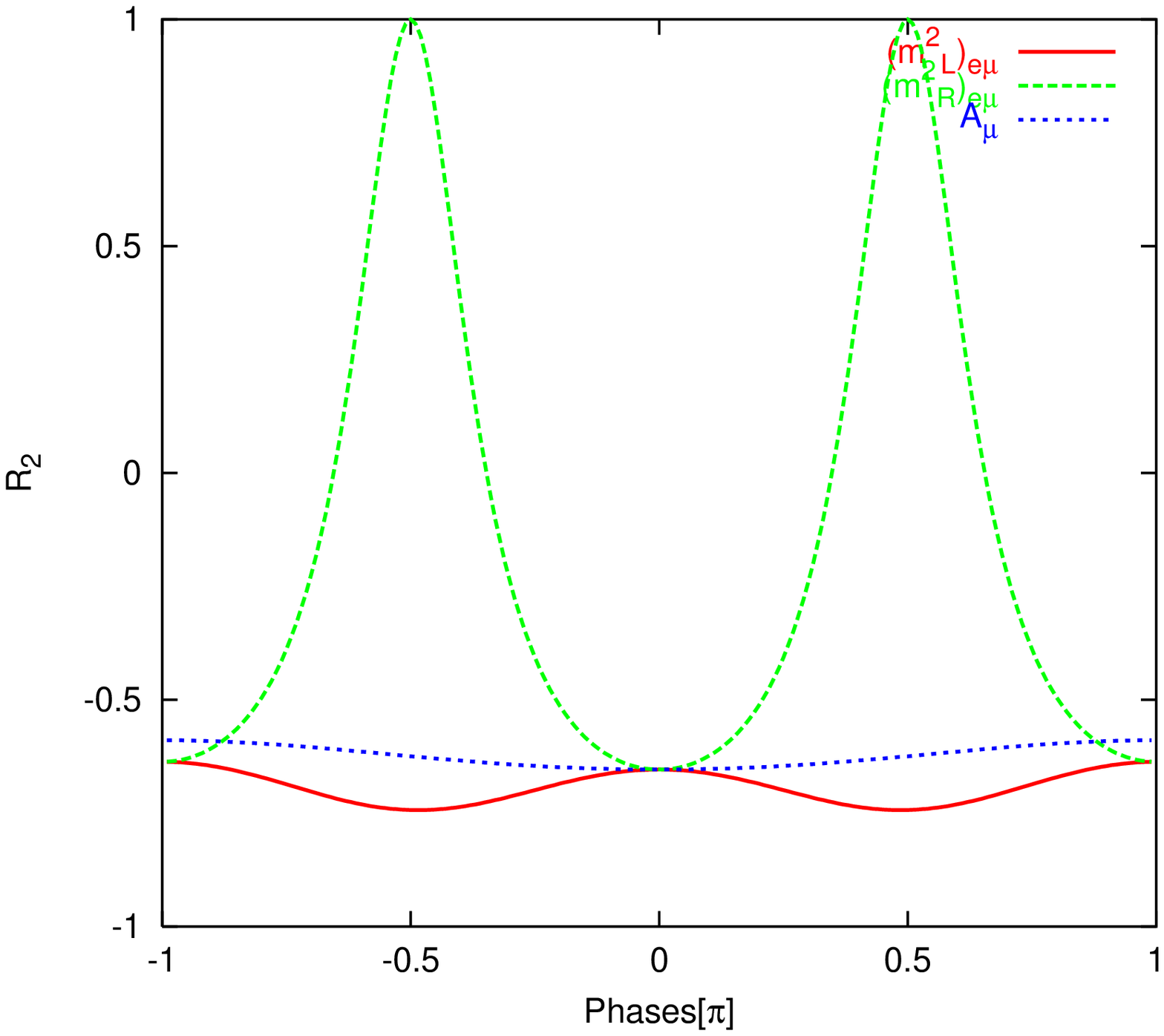}\hspace{5mm}\includegraphics[scale=0.4]{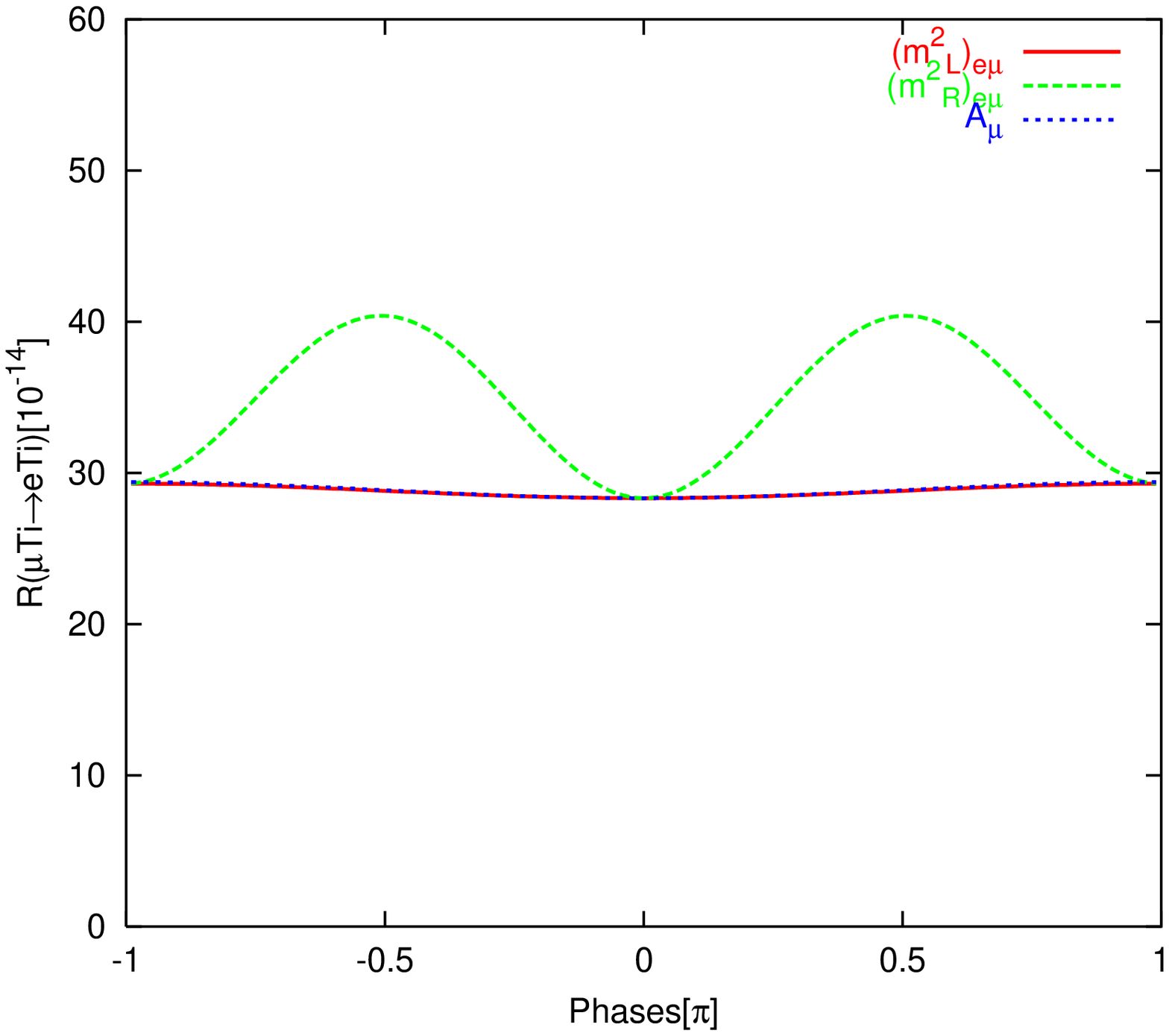}}
\centerline{\hspace{1.2cm}(c)\hspace{7cm}(d)}
\centerline{\vspace{-1.5cm}}
\end{center}
\caption{ Observable quantities in the $\mu- e$ conversion
experiment versus the phases of $A_\mu$, $(m^2_L)_{e\mu }$ and
$(m^2_R)_{e\mu }$. The vertical axes  in Figs. (a)-(d) are
respectively
  $\overline{\langle s_{T_1}}\rangle$, $\overline{\langle
s_{T_2}}\rangle$, $R_2$  and $R(\mu {\rm Ti}\rightarrow e {\rm
Ti})$. The input parameters correspond to the $P3$ benchmark
proposed in \cite{Heinemeyer:2007cn}: $|\mu|=400$~GeV,
$m_0=1000$~GeV, ${\rm M}_{1/2}=500$~GeV and $\tan \beta=10$ and we
have set $|A_{\mu}|$=$|A_{e}|$=700~GeV. All the LFV elements of the
slepton mass matrix are set to zero except $(m_L^2)_{e \mu}=250~{\rm
GeV}^2$ and $(m_R^2)_{e \mu}=12500~{\rm GeV}^2$.  We have taken
$\mathbb{P}_\mu=20 \%$.}\label{hier-conversion}
\end{figure}

\begin{figure}
\begin{center}
\centerline{\vspace{-1.2cm}}
\centerline{\includegraphics[scale=0.4]{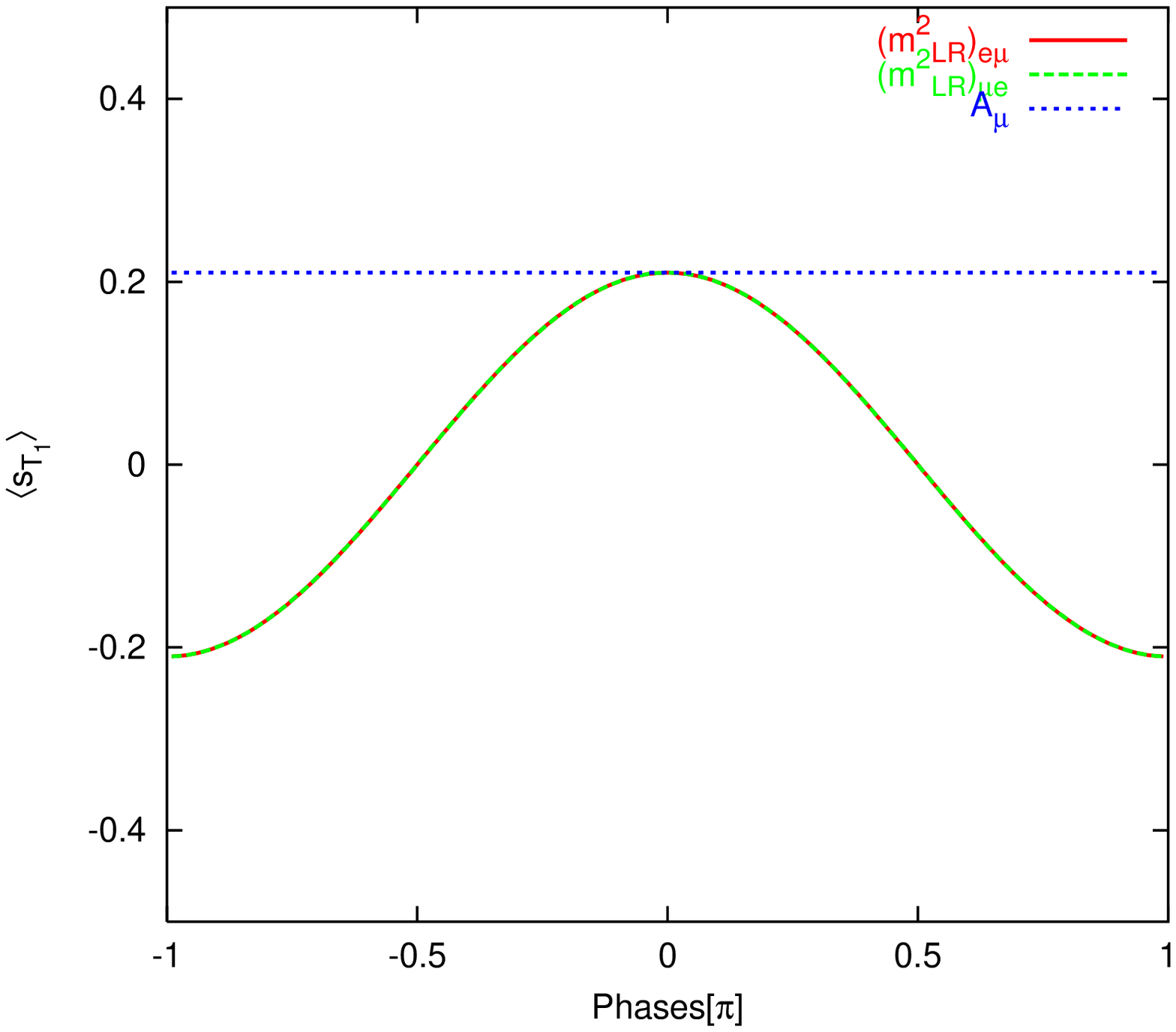}\hspace{5mm}\includegraphics[scale=0.4]{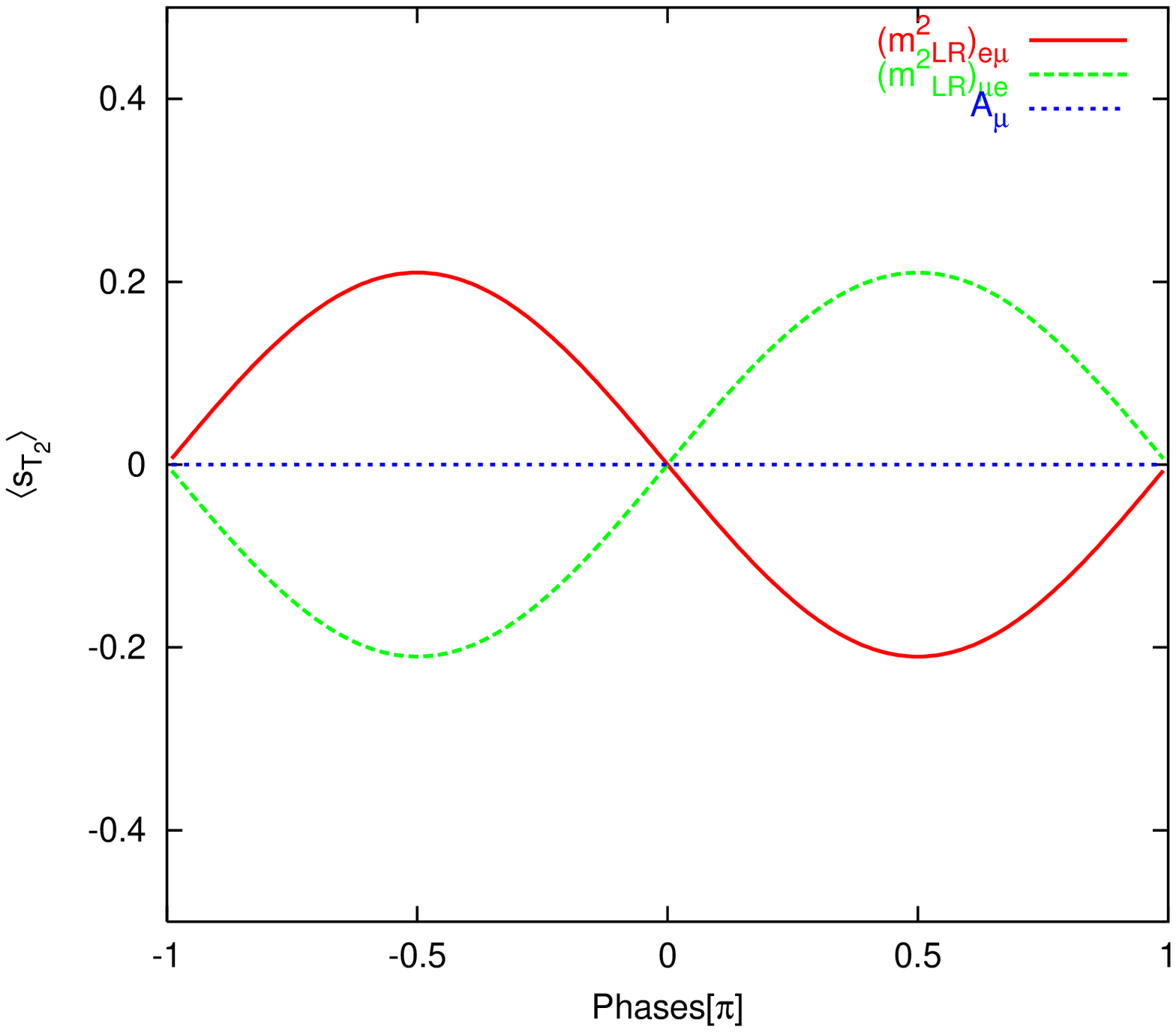}}
\centerline{\hspace{1.2cm}(a)\hspace{7cm}(b)}
\centerline{\includegraphics[scale=0.4]{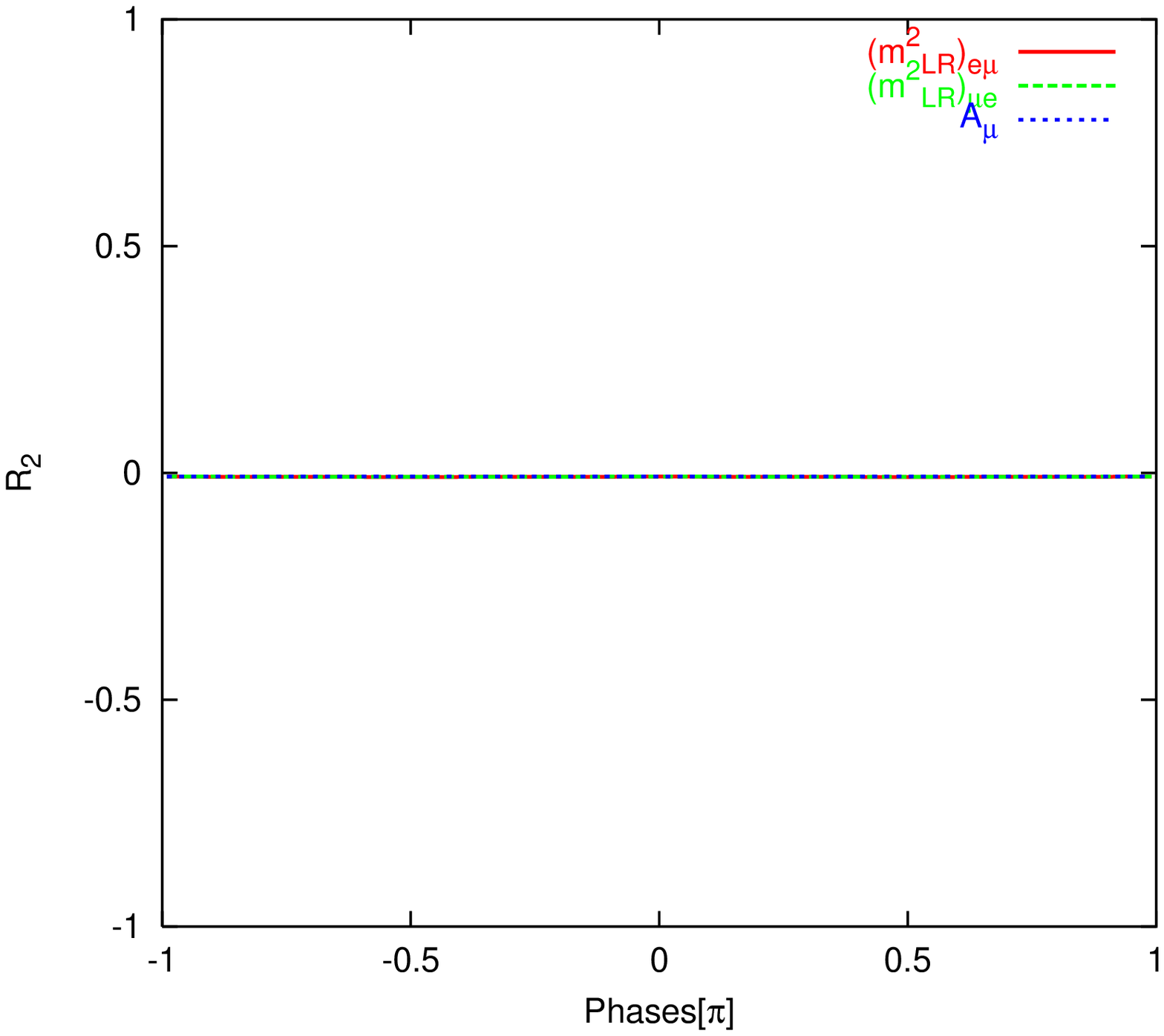}\hspace{5mm}\includegraphics[scale=0.4]{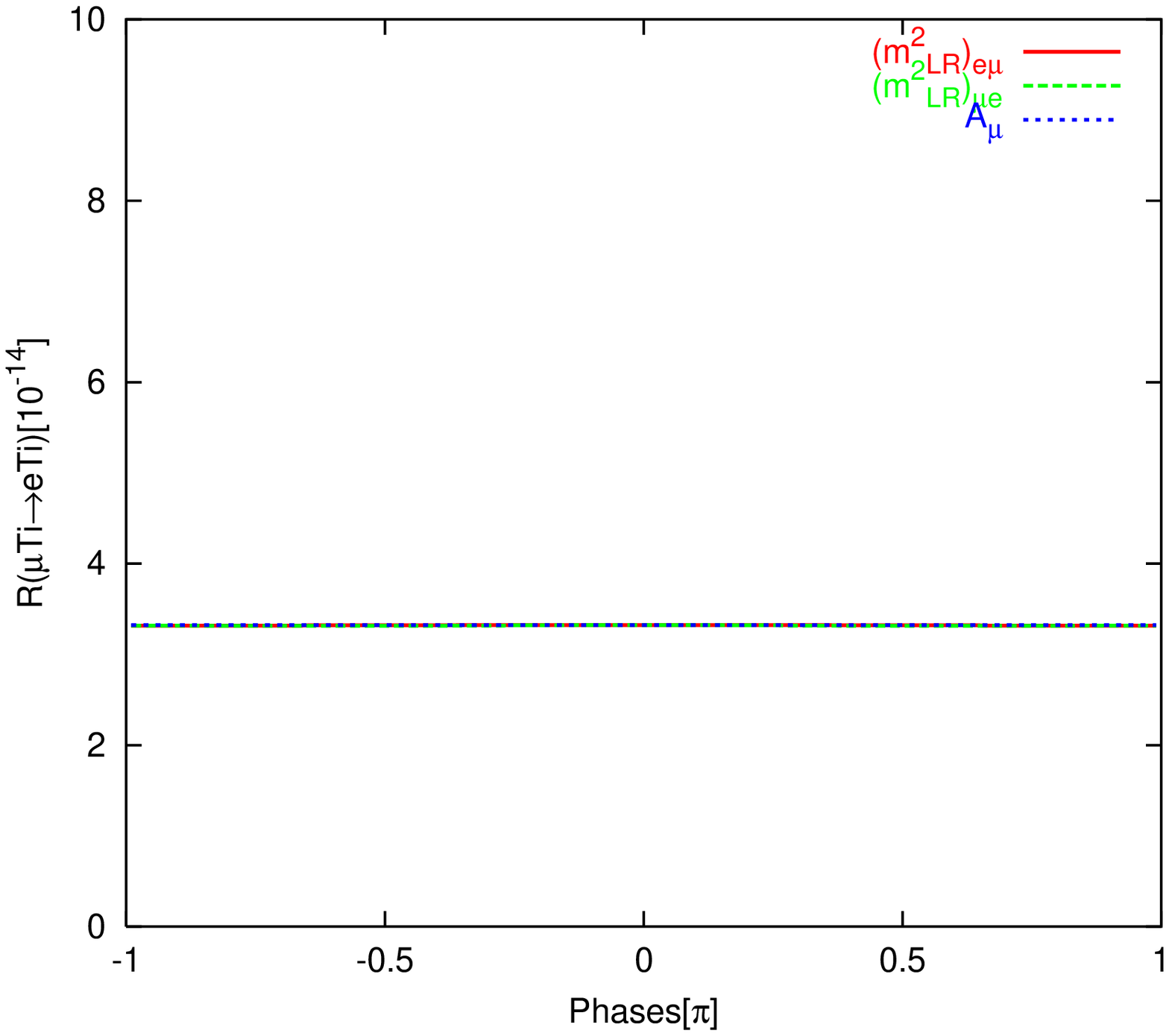}}
\centerline{\hspace{1.2cm}(c)\hspace{7cm}(d)}
\centerline{\vspace{-1.5cm}}
\end{center}
\caption{Observable quantities in the $\mu- e$ conversion
experiment versus the phases of $A_\mu$, $(m^2_{LR})_{e\mu }$ and
$(m^2_{LR})_{\mu e}$. The vertical axes  in Figs. (a)-(d) are
respectively $\overline{\langle s_{T_1}}\rangle$,
$\overline{\langle s_{T_2}}\rangle$, $R_2$  and $R(\mu {\rm
Ti}\rightarrow e {\rm Ti})$.   The input parameters correspond to
the $P3$ benchmark proposed in \cite{Heinemeyer:2007cn}:
$|\mu|=400$~GeV, $m_0=1000$~GeV, ${\rm M}_{1/2}=500$~GeV and $\tan
\beta=10$ and we have set $A_{\mu}$=$A_{e}$=700~GeV. All the LFV
elements of the slepton mass matrix are set to zero except
$(m_{LR}^2)_{e\mu}(=A_{e\mu}\langle H_d\rangle)$=$14~{\rm GeV}^2$
and $(m_{LR}^2)_{\mu e}(=A_{\mu e}\langle H_d\rangle)$=$14~{\rm
GeV}^2$.  We have taken $\mathbb{P}_\mu=20
\%$.}\label{LRconversion}
\end{figure}

The CP-violating phases that can in principle show up in the
polarizations studied in the previous sections are the phases of
$A_i$,  the $\mu$-term, $M_1$ (the Bino mass) and phases of LFV
elements of mass matrices in soft supersymmetry breaking
Lagrangian. The strong bound on the electric dipole moment of the
electron implies  strong bounds on the phases of $A_e$, $\mu$ and
$M_1$ ({\it see}, however \cite{first}). For this reason, in this
paper, we  set the phases of these parameters equal to zero and
focus on the effects of the phases of $A_\mu$
 and the LFV elements of mass
matrices. In the present analysis, we focus on the effects of the
$e\mu$ elements. Effects of $e\tau$  and $\mu \tau$ elements will
be explored elsewhere.

Once we turn on the LFV terms, the phase of $A_\mu$ as well as the
phases of the LFV elements can contribute to $d_e$ at one loop
level \cite{main,Bartl}. We therefore have to make sure that the
bounds on $d_e$ are satisfied. For the parameters that we have
considered in this analysis, the contributions of the phases of
${e \mu}$ elements to  $d_e$ are of order of $\sim 10^{-29}~e~{\rm
cm}$ and well below the present bound \cite{pdg}. The contribution
of the phase of $A_\mu$ is even lower by one order of magnitude.
In the next section, we shall discuss the role of the forthcoming
results of $d_e$ searches in reducing the degeneracies.

 As the
reference point, we have chosen the mass spectra corresponding to
the $P3$ benchmark which has been proposed in
\cite{Heinemeyer:2007cn}. We have however let $A_e$ and $A_\mu$
deviate
 from the corresponding  values at
the benchmark $P3$.
%Varying $A_\tau$ would have dramatic effects
%on the dark matter density  at parts of the parameter space for
%which the coannihilation between $\tilde{\tau}_1$ and
%$\tilde{\chi}_1^0$ in the early universe determines the dark
%matter density; {\it i.e.,} for benchmarks at which
%$m_{\tilde{\tau}_1} \approx m_{\tilde{\chi}^0_1}$.  This is not
%the case for the $P3$ benchmark.
The values of $A_i$ and $A_{ij}$ are chosen such that they satisfy
the constraints from  Color and Charge Breaking (CCB) as well as
Unbounded From Below (UFB) considerations \cite{UFB}. The rest of
the bounds and restrictions on  the parameters of supersymmetry
are undisturbed by  varying $A_i$.

Figs. (\ref{sim-muegamma}-\ref{LRdecay}) shows $R_1$, Br$(\mu \to
e \gamma)$, $\overline{\langle P_{T_1}s_{T_1}\rangle}$ and
$\overline{\langle P_{T_1}s_{T_2}\rangle}$ ({\it see},
Eqs.~(\ref{R1},\ref{average-theta},\ref{average-theta2}) for
definitions) versus the phases of $A_\mu$ and the LFV elements. We
have set $|A_e|=|A_\mu|$ however the results  are robust against
varying  the values of $|A_e|$  as expected. In Fig.
(\ref{sim-muegamma}), we have taken $A_{ij}$ and all the LFV
elements of the slepton mass matrix other than $(m^2_L)_{e\mu }$
and $(m^2_R)_{e\mu }$  equal to zero. Notice that
 $(m^2_L)_{e\mu }$ and $(m^2_R)_{e\mu }$ have been chosen such
that $Br(\mu\rightarrow e\gamma)$ lies close to its present
experimental upper bound. As seen from
Fig.~(\ref{sim-muegamma}-c), for such choice of $(m^2_L)_{e\mu }$
and $(m^2_R)_{e\mu }$, $R_1$ is close to zero which means
$|A_L|\approx |A_R|$. As a result, we expect the transverse
polarization to be sizable.
Figs.~(\ref{sim-muegamma}-a,\ref{sim-muegamma}-b) demonstrate that
this expectation is fulfilled. From
Figs.~(\ref{sim-muegamma}-a,\ref{sim-muegamma}-b), we also observe
that the sensitivity of the transverse polarization to the phases
of $(m^2_L)_{e\mu }$ and $(m^2_R)_{e\mu }$ is significant so by
measuring  these polarizations with a moderate accuracy one can
extract information on these phases. However at this benchmark,
the sensitivity to the phase of $A_\mu$ is quite low.

 The input of
Fig.~(\ref{hier-muegamma}) is similar to that of
Fig.~(\ref{sim-muegamma}) except that a hierarchy is assumed
between the left and right LFV elements: $|(m^2_L)_{e\mu}|\ll
|(m^2_R)_{e \mu}|$. As expected in this case, $R_1\approx 1$ and
the transverse polarizations are small. To draw
Fig.~(\ref{LRdecay}), we have set the LFV elements of $m_L^2$ and
$m_R^2$ equal to zero and instead we have set $A_{e \mu},A_{\mu
e}\ne 0$. As seen in Fig.~(\ref{LRdecay}) in this case, the
transverse polarizations can be sizeable.

Figs. (\ref{sim-conversion}-\ref{LRconversion}) show $R_2$, $R(\mu+
Ti \to e +Ti)$, $\overline{\langle s_{T_1}\rangle}$ and
$\overline{\langle s_{T_2}\rangle}$ ({\it see},
Eqs.~(\ref{R2},\ref{mean-conversion-sT1},\ref{mean-conversion-sT2})
for definitions) versus the phases of $A_\mu$ and the LFV elements.
To draw the figures corresponding to the $\mu-e$ conversion, we have
taken $\mathbb{P}_\mu =20 \%$. If the technical difficulties of
polarizing the muon in the $\mu-e$ conversion experiment is overcome
and higher values of $\mathbb{P}_\mu$ is achieved,
$\overline{\langle s_{T_1}\rangle}$ and $\overline{\langle
s_{T_2}\rangle}$ can become larger. Obviously, for a given value of
$\mathbb{P}_\mu$, $\overline{\langle s_{T_1}\rangle}$ and
$\overline{\langle s_{T_2}\rangle}$  have to be re-scaled by
$(\mathbb{P}_\mu/20 \%)$. Apart from the polarization, the input
parameters in
Figs.~(\ref{sim-conversion},\ref{hier-conversion},\ref{LRconversion})
are respectively the same as the input parameters in
Figs.~(\ref{sim-muegamma},\ref{hier-muegamma},\ref{LRdecay}). Notice
that in this case, too, the sensitivity to the phase of $A_\mu$ is
low.  From Fig.~(\ref{hier-muegamma}), we observe that
$|\overline{\langle s_{T_2}\rangle}|$ increases  more rapidly with
$\sin[\arg [(m_L^2)_{e \mu}]]$  than with $\sin[\arg
[(m_R^2)_{e\mu}]]$. For $|(m_L^2)_{e \mu}|\ll |(m_R^2)_{e\mu}|$
cases, at first sight, higher sensitivity to $\arg[(m_L^2)_{e\mu}]$
may sound counterintuitive. However, notice that as
$|\sin(\arg[(m_R^2)_{e\mu}])|$ increases, $R_2$ rapidly converges to
one which means $K_R\gg K_L$ and therefore $\langle s_{T_2}\rangle
\propto {\rm Im}[K_RK_L^*]/(|K_L|^2+|K_R|^2) \to 0.$

\begin{figure}
\begin{center}
\centerline{\vspace{-1.2cm}}
\centerline{\includegraphics[scale=0.4]{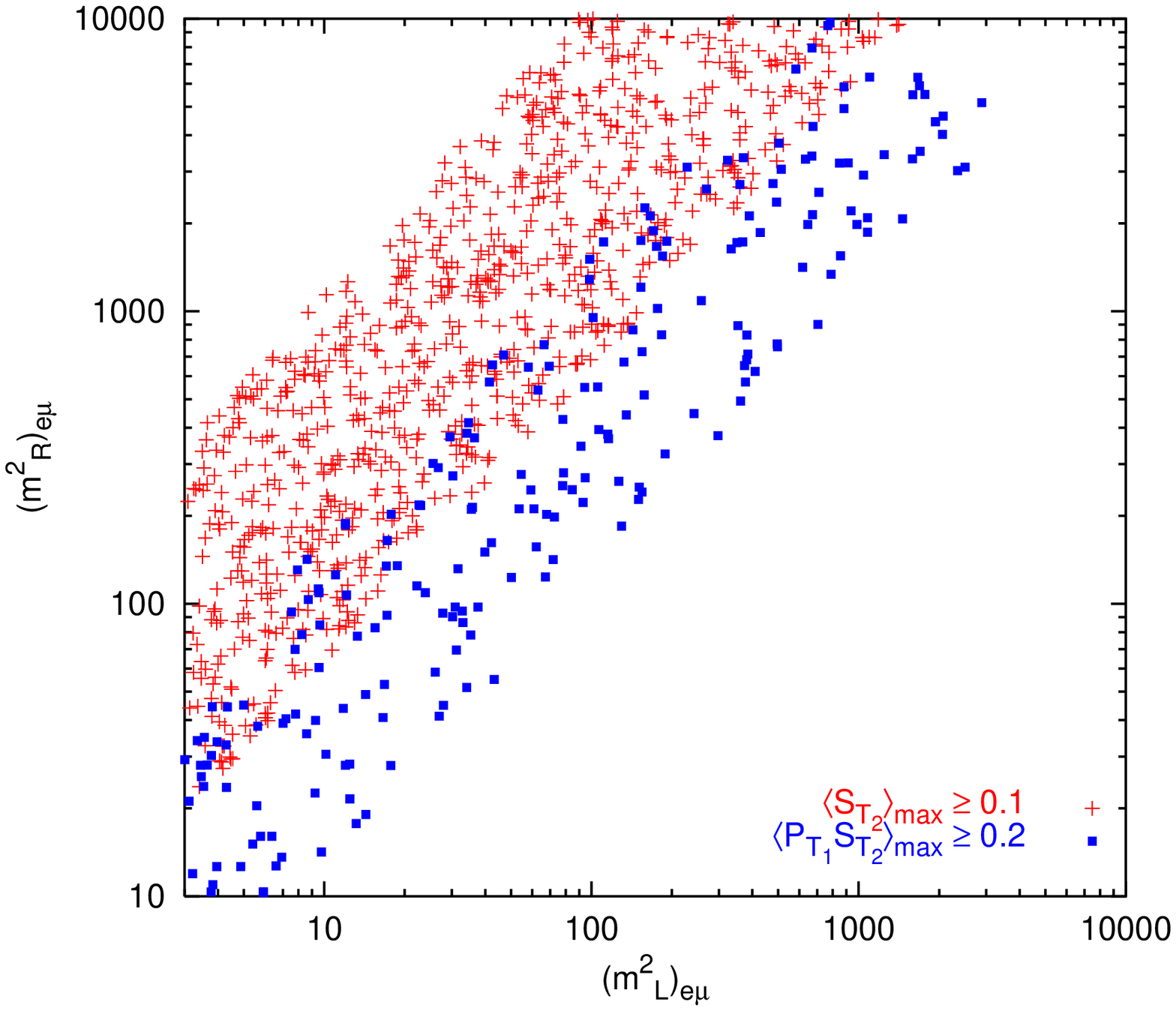}\hspace{5mm}\includegraphics[scale=0.4]{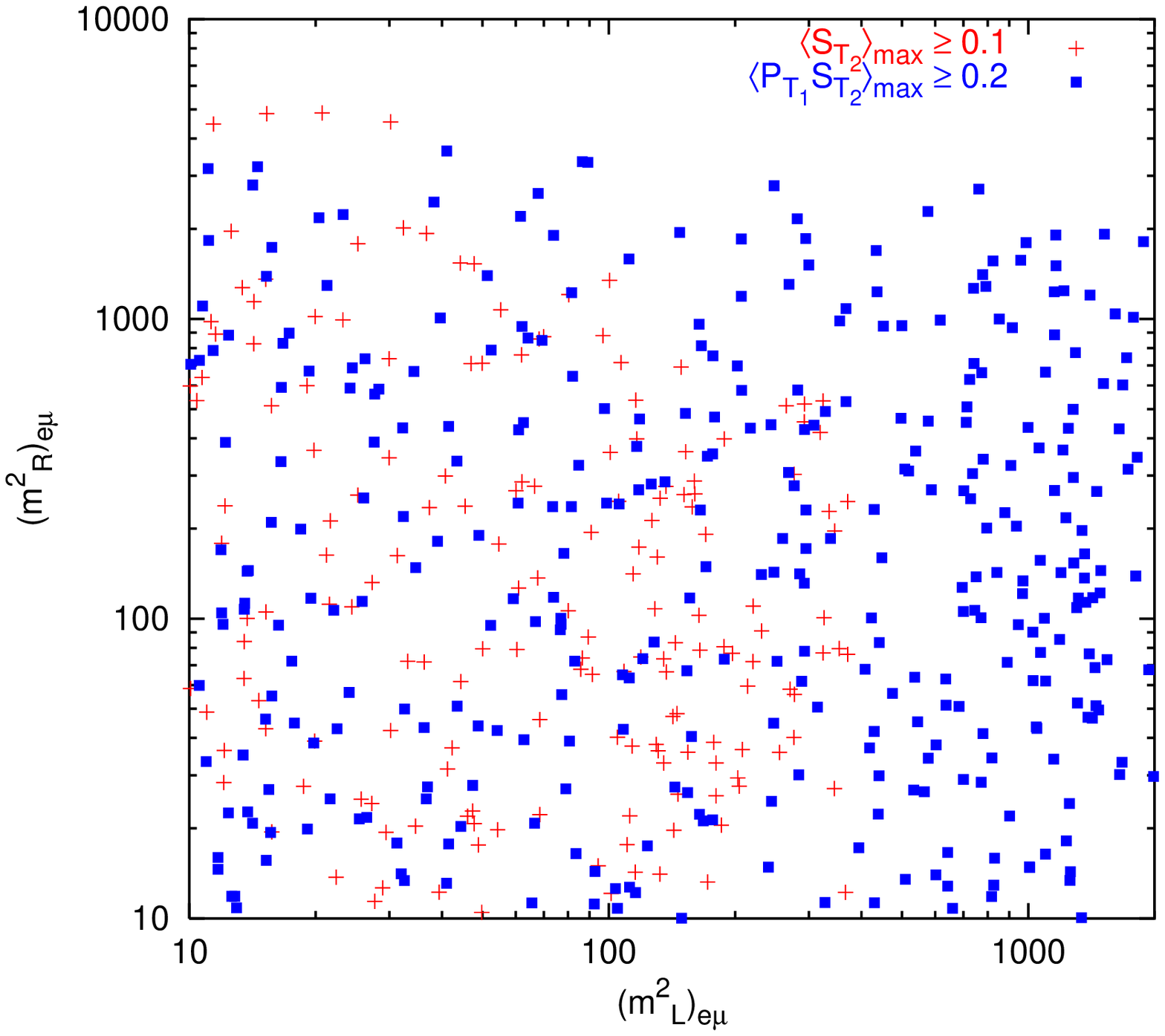}}
\centerline{\hspace{1.2cm}(a)\hspace{7cm}(b)}
\centerline{\includegraphics[scale=0.4]{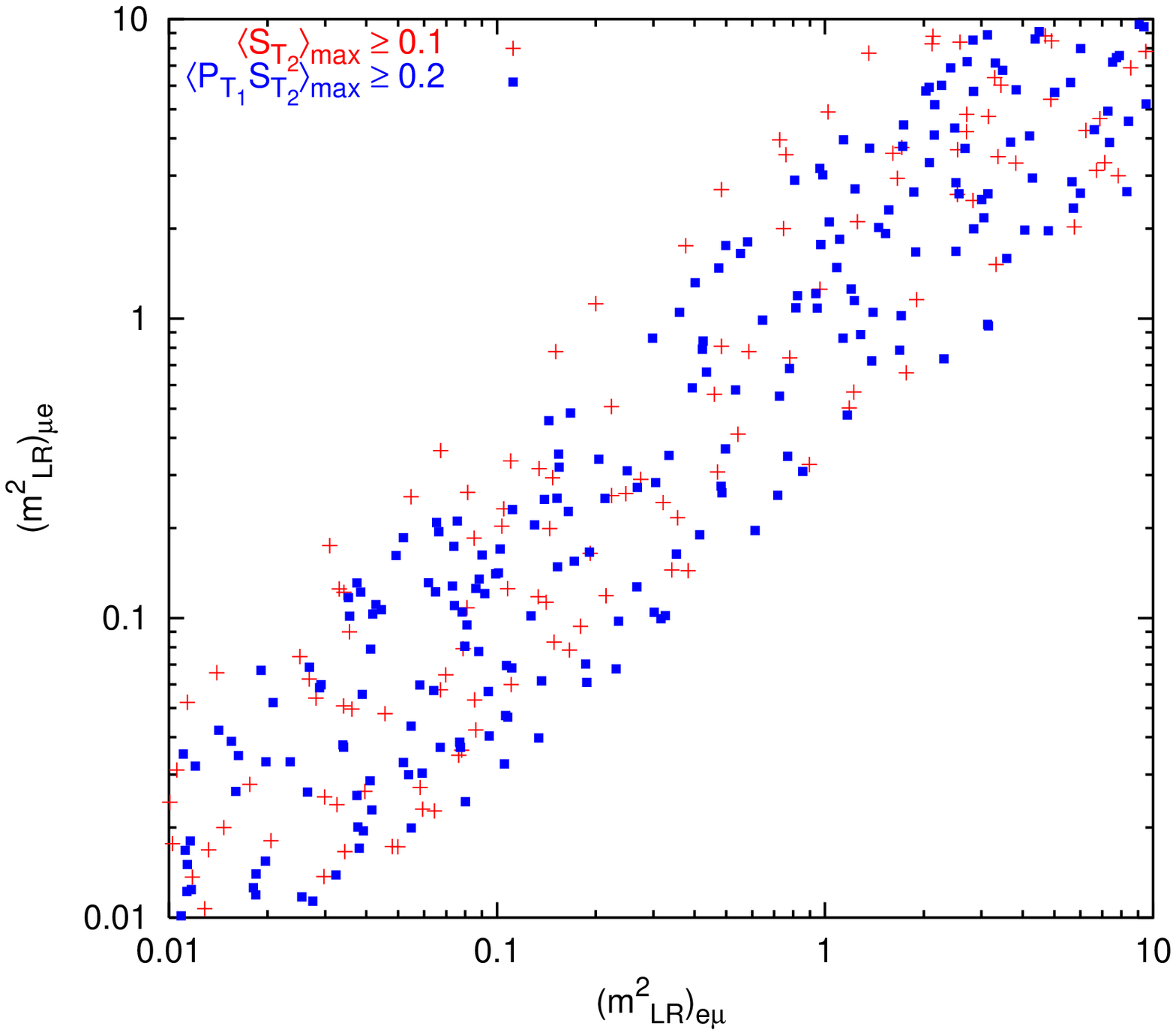}\hspace{5mm}\includegraphics[scale=0.4]{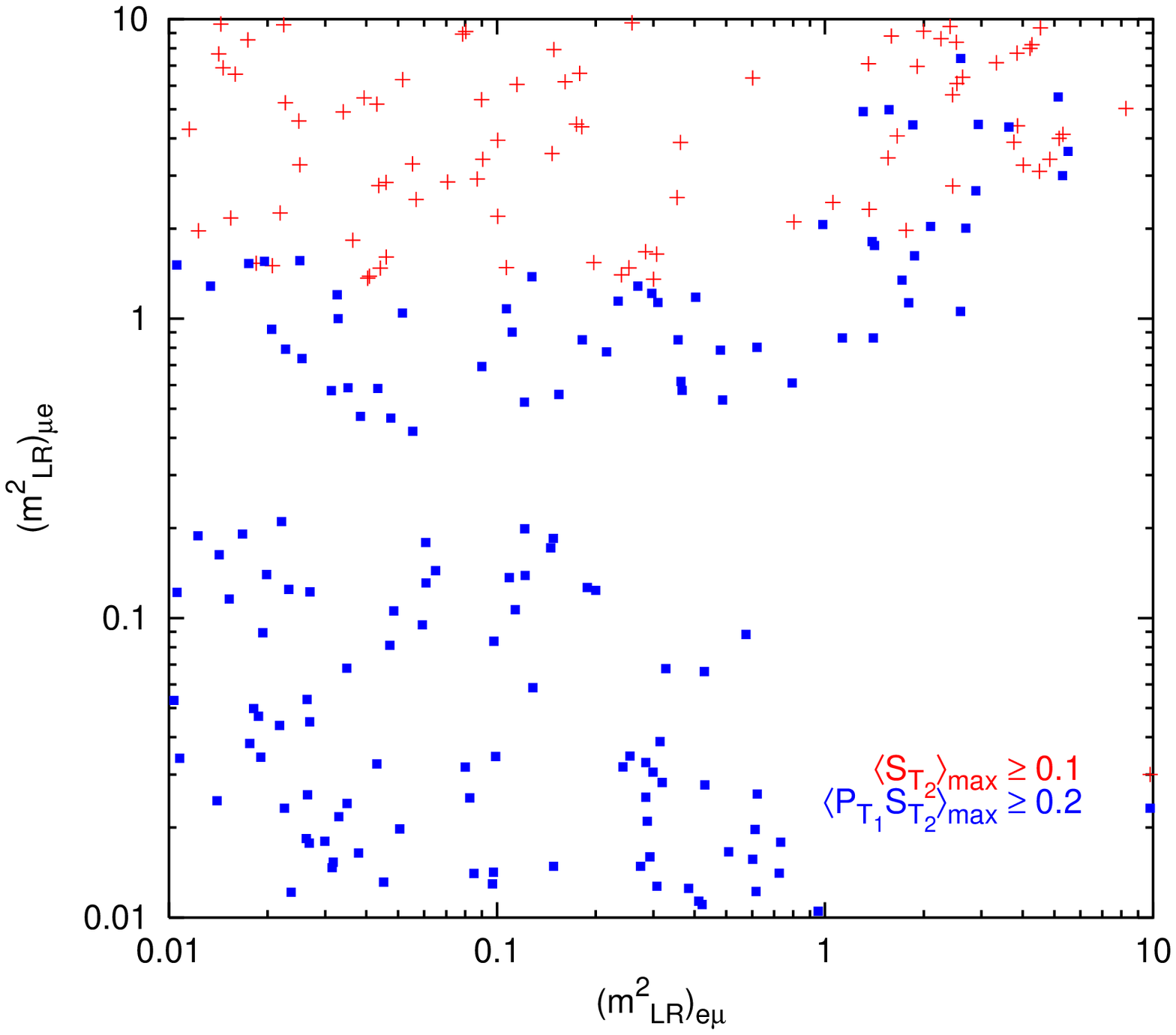}}
\centerline{\hspace{1.2cm}(c)\hspace{7cm}(d)}
\centerline{\vspace{-1.5cm}}
\end{center}
\caption{ Scatter plots showing points for which
$\overline{\langle s_{T_2}\rangle}$ at the $\mu N\to e N$
experiment with  $\mathbb{P}_\mu=20\%$ and $\overline{\langle
P_{T_1}s_{T_2}\rangle}$ at the $\mu \to e \gamma$ experiment with
$\mathbb{P}_\mu=100\%$ are sizeable. The points depicted by plus
(square)  show the points at which the maximum value of
$\overline{\langle s_{T_2}\rangle}$  ($\overline{\langle
P_{T_1}s_{T_2}\rangle}$) is larger than 0.1 (0.2). The input for
LF conserving parameters are the same as the input in
Fig.~\ref{sim-muegamma}: {\it i.e.,} the P3 benchmark with
$A_{\mu}=A_{e}$=700~GeV. In Fig.~(a) all the LFV elements of the
slepton mass matrix are set to zero except $(m_L^2)_{e \mu}$ and
$(m_R^2)_{e \mu}$ which are randomly chosen respectively from
($3~{\rm GeV}^{2},3 \times 10^3~{\rm GeV}^2)$ and $(10~{\rm
GeV}^2,10^4~{\rm GeV}^2 )$ at a logarithmic scale. The maximum
polarization correspond to $\arg[(m_L^2)_{e\mu}]=\pi/2$ and
$\arg[(m_R^2)_{e\mu}]=0$. Fig.~(b) is similar to Fig.~(a) except
that $(m^2_{LR})_{e\mu}$=$(m^2_{LR})_{\mu e}$=$4~{\rm GeV}^{2}$
and $(m_L^2)_{e \mu}$ and $(m_R^2)_{e \mu}$ are  chosen
respectively from ($2~{\rm GeV}^{2},2 \times 10^3~{\rm GeV}^2)$
and $(5~{\rm GeV}^2,5\times10^3~{\rm GeV}^2 )$. In Fig.~(c), we
have set $(m^2_{L})_{e\mu }$=$(m^2_{R})_{\mu e}$=$0$ and allowed
$(m^2_{LR})_{e\mu }$ and $(m^2_{LR})_{\mu e}$ to pick up random
values at a logarithmic scale from the interval $(0.01~{\rm
GeV}^2,10~{\rm GeV}^2)$. In Fig.~(d), we have set $(m^2_{L})_{e\mu
}$=$100~{\rm GeV}^2$, $(m^2_{R})_{e\mu }$=$400~{\rm GeV}^2$ and
allowed $(m^2_{LR})_{e\mu }$ and $(m^2_{LR})_{\mu e}$ to pick up
random values  from the interval $(0.01~{\rm GeV}^2,10~{\rm
GeV}^2)$. }\label{robustness}
\end{figure}

It is remarkable that in the case  of Fig.~(\ref{sim-conversion})
for which $(m_L^2)_{e \mu}\sim (m^2_R)_{e\mu}$, $R_2$ is close to
one and the transverse polarizations is relatively small but in the
case of Fig.~(\ref{hier-conversion})   for which $(m_R^2)_{e \mu}=50
(m_L^2)_{e \mu}$, $(\left|1-|R_2|\right| \sim 1)$ and the transverse
polarizations become sizeable. We have explored higher hierarchy
between the left and right LFV elements and have found that for
$(m_L^2)_{e\mu}\stackrel{<}{\sim} 500 (m_R^2)_{e\mu}$, $\langle
s_{T_1}\rangle$ and $\langle s_{T_2} \rangle$ diminish. Contrasting
Figs.~(\ref{sim-conversion},\ref{hier-conversion}) with
Figs.~(\ref{sim-muegamma},\ref{hier-muegamma}), we find that the
polarization studies at the $\mu \to e \gamma$ and $\mu-e$
conversion experiments can be complementary. That is if $1-|R_1|\sim
{\rm few} \times 0.01$, transverse polarization in the $\mu \to e
\gamma$ will become small making the derivation of the CP-violating
phases more challenging. However  there is still the hope to derive
the phases by polarization studies at the $\mu-e$ conversion
experiments. We shall discuss this point in more detail in the
description of Fig.~\ref{robustness}.

Notice that in Figs.~(\ref{sim-muegamma}-\ref{LRconversion}),
which all correspond to the benchmark $P_3$, sensitivity to the
phase of $A_\mu$ is low. This is expected because the effect of
$A_\mu$ is suppressed by $\tan \beta=10$. We have checked for the
robustness of this result and found that for most of the parameter
space with large $\tan \beta$, sensitivity to the phase of $A_\mu$
is low but there are points at which sensitivity to $\phi_{A_\mu}
$ is considerable; {\it e.g.,} at $\delta$ benchmark which has
been proposed in \cite{delta}.

 The following remarks are in order:
\begin{itemize}

\item

In all of these sets of diagrams, maximal $|\overline{\langle
s_{T_1}\rangle} |$ corresponds to $|\overline{\langle
s_{T_2}\rangle}|=0$ and vice versa. This is expected from Eqs.
(\ref{mean-conversion-sT1}) and (\ref{mean-conversion-sT2}) because
$\overline{\langle s_{T_1}\rangle}$ and $ \overline{\langle
s_{T_2}\rangle}$ are respectively given by the real and imaginary
parts of the same combinations. For general values of the phases,
$|\overline{ \langle s_{T_1} \rangle}|^2+|\overline{ \langle s_{T_2}
\rangle}|^2$ is solely given by the absolute values of $K_L$ and
$K_R$, and is independent of their relative phase. Remember that
$|K_R|$ and $|K_L|$ can be extracted by studying the angular
distribution  of the electron without measuring its spin. Thus, the
simultaneous measurement of $R_2$, $\overline{\langle s_{T_1}
\rangle}$ and $\overline{\langle s_{T_2} \rangle}$ provides a
cross-check. A similar consideration holds  for $R_1$,
$\overline{\langle P_{T_1} s_{T_1}\rangle }$ and $ \overline{\langle
P_{T_1} s_{T_2}\rangle }$, too.

\item
When all the phases are set   equal to zero, $\langle s_{T_2}
\rangle$ and $\langle P_{T_1}s_{T_2}\rangle$
 vanish but $\langle s_{T_1} \rangle$ and $\langle
P_{T_1}s_{T_1}\rangle$ can be nonzero. Thus, for the purpose of
establishing CP, it will be more convenient to measure $\langle
s_{T_2} \rangle$ or $\langle P_{T_1}s_{T_2}\rangle$.  This is
expected from
Eqs.~(\ref{average-theta},\ref{average-theta2},\ref{mean-conversion-sT1},\ref{mean-conversion-sT2}).

\item
When $(m_{LR}^2)_{e\mu}=(m_{LR}^2)_{\mu e}=0$, in the case of $\mu
\to e \gamma$, there is a symmetry under $\arg[(m_L^2)_{e\mu }]
\leftrightarrow -\arg[ (m_R^2)_{e\mu }]$ [{\it see}
Figs.~(\ref{sim-muegamma},\ref{hier-muegamma})] but in the case of
the $\mu-e$ conversion, there is not such a symmetry [{\it see}
Figs.~(\ref{sim-conversion},\ref{hier-conversion})]. Moreover,
while the dependence of $R_1$ on the phases is very mild, $R_2$
can dramatically change with varying some of the phases (see, {\it
e.g.,} Fig.~(\ref{hier-conversion}-c)). This can be better
understood in the limit of the LFV mass insertion approximation.
Remember that observables in the $\mu\to e\gamma$ decay are given
by $A_L$ and $A_R$ for $(m_{LR}^2)_{\mu e}=(m_{LR}^2)_{e\mu}=0$.
To leading approximation, $A_L$ and $A_R$ are respectively
proportional to $(m_R^2)_{e\mu }$ and $(m_L^2)_{e\mu }$.  As a
result, when we vary the phase of $(m^2_R)_{e\mu }$, only the
phase of $A_L$ changes. Similarly varying $\arg[(m_L^2)_{e\mu }]$
only changes $\arg[A_R]$. Since $R_1$ depends only on the absolute
values of $A_R$ and $A_L$, it should not change with varying the
phases. Remember that $\langle P_{T_1} s_{T_1} \rangle$ and
$\langle P_{T_1} s_{T_2} \rangle$ are given by ${\rm Re}[A_L
A_R^*]$ and ${\rm Im}[A_L A_R^*]$ which to leading order are
proportional to ${\rm Re}[(m_R^2)_{e\mu} (m_L^2)_{e\mu}^*]$ and
${\rm Im}[(m_R^2)_{e\mu} (m_L^2)_{e\mu}^*]$. Thus, there should be
a symmetry under $\arg[(m_L^2)_{e\mu }] \leftrightarrow
-\arg[(m_R^2)_{e\mu }]$ for $(m_{LR}^2)_{\mu
e}=(m_{LR}^2)_{e\mu}=0$. Observables in the $\mu-e$ conversion
case depend on $K_L$ and $K_R$. Unlike $A_L$ and $A_R$, each of
$K_L$ and $K_R$ can receive contributions from both $(m_L^2)_{e
\mu}$ and $(m_R^2)_{e \mu}$. Thus, the above argument does not
apply here. Similar consideration holds for the case that
$(m_{LR}^2)_{\mu e}$ and $(m_{LR}^2)_{e\mu}$ are nonzero but
$(m_{R}^2)_{e\mu }=(m_{L}^2)_{e\mu}=0$ ({\it see},
Figs.~\ref{LRdecay} and \ref{LRconversion}).  As expected, when
$(m_R^2)_{e\mu}$, $(m_L^2)_{e \mu}$, $(m_{LR}^2)_{\mu e}$ and
$(m_{LR}^2)_{e\mu}$ are all nonzero, the symmetries under
 $\arg[(m_L^2)_{e\mu }]
\leftrightarrow -\arg[ (m_R^2)_{e\mu }]$ and $\arg[(m_{LR}^2)_{\mu
e}] \leftrightarrow -\arg[ (m_{LR}^2)_{ e\mu}]$ disappear.

\item In this analysis, we have considered the $\mu-e$ conversion
only on Titanium. It is possible to perform the experiment on
other  nuclei such as Au and Al, too.  From
Eqs.~(\ref{partial-conversion},\ref{aANDb}), we find that the
effects change with changing the nuclei (with change of $N$ and
$Z$). In principle, by studying the conversion rate on different
nuclei, one can derive information on different combinations of
the phases. However, in practice since the ratio $N/Z$ for
different nuclei in question are more or less the same (the
difference between $N/Z$ of Au and Al is about $20\%$), $\langle
s_{T_1}\rangle$, $\langle s_{T_2}\rangle$ and $R_2$ for different
nuclei turn out to be close to each other. Only if $\langle
s_{T_i}\rangle$ can be measured with accuracy better than 5\%
({\it i.e.,}  $\delta \langle s_{T_i}\rangle/\langle
s_{T_i}\rangle<5 \%$), using different nuclei will help us to
solve degeneracies.

\end{itemize}

Scatter plots shown in Fig.~\ref{robustness} demonstrate the
configurations of the LFV elements  where $\overline{\langle
s_{T_2}\rangle}$ or $\overline{\langle P_{T_1}s_{T_2}\rangle}$ can
be sizeable. That is where maximal values of $\langle
s_{T_2}\rangle$ and $\langle P_{T_1}s_{T_2}\rangle$ are
respectively larger than $0.1$ and $0.2$. In Fig. (a) and (c)
where only a pair of LFV are nonzero, only within a band $\langle
s_{T_2}\rangle$ and $\langle P_{T_1}s_{T_2}\rangle$ can be large.
This is expected because when there is a hierarchy between the
nonzero elements, we expect a hierarchy between $K_L$ and $K_R$ as
well as between $A_L$ and $A_R$ thus $\langle s_{T_2}\rangle$ and
$\langle P_{T_1}s_{T_2}\rangle$ are suppressed. In Figs.~(b) and
(d), $(m_L^2)_{e\mu}$, $(m_R^2)_{e\mu}$, $(m_{LR}^2)_{e\mu}$ and
$(m_{LR}^2)_{\mu e}$ are all nonzero. Notice that depending on the
configuration of the LFV elements,
 the  regions over   which $\langle
s_{T_2}\rangle$ and $\langle P_{T_1} s_{T_2}\rangle$ are large can
have partial (like Figs.~a and d) or complete (like Figs.~b and
c).
 This confirms our observation regarding the previous figures.
 In the case of overlap, one can employ both experiments to derive information on the
 CP-violating phases. In the
next section, we discuss how by combining the information from
these two experiments, one can derive extra information and
resolve degeneracies.

\section{Resolving Degeneracies \label{degeneracies}}
As discussed in the previous sections, all the observables in the
$\mu \to e \gamma$ experiment are determined by a pair of effective
couplings ($A_L,A_R$) which in turn  receive contributions from
various parameters in the underlying theory. By measuring Br($\mu
\to e \gamma$), $R_1$ and either of $\overline{\langle
P_{T_1}s_{T_1}\rangle}$ and $\overline{\langle P_{T_1}
s_{T_2}\rangle}$ ({\it see},
Eqs.~(\ref{average-theta},\ref{average-theta2})), one can
reconstruct both $A_L$ and $A_R$ (up to a common phase). However,
because of the degeneracies, it is not possible to unambiguously
derive the values of the LFV elements and the CP-violating phases of
the underlying theory from $A_L$ and $A_R$.

Similarly to the $\mu \to e \gamma$ experiment, the observable
quantities in the $\mu N \to e N$ experiment are given by a pair
of parameters ($K_L,K_R$) which depend on the LFV masses and
CP-violating phases of the underlying theory. By measuring $R(\mu
N\to eN)$, $R_2$ and  either of $\overline{\langle
s_{T_1}\rangle}$ and $\overline{\langle s_{T_2}\rangle}$ ({\it
see}, Eqs. (\ref{mean-conversion-sT1},\ref{mean-conversion-sT2})),
it is possible to reconstruct $|K_L|$, $|K_R|$ and their relative
phase; however, deriving the LFV and CP-violating parameters of
the underlying theory from $(K_L,K_R)$ would suffer from
degeneracies.
\begin{figure}
\begin{center}
\centerline{\vspace{-1.2cm}}
\centerline{\includegraphics[scale=0.5]{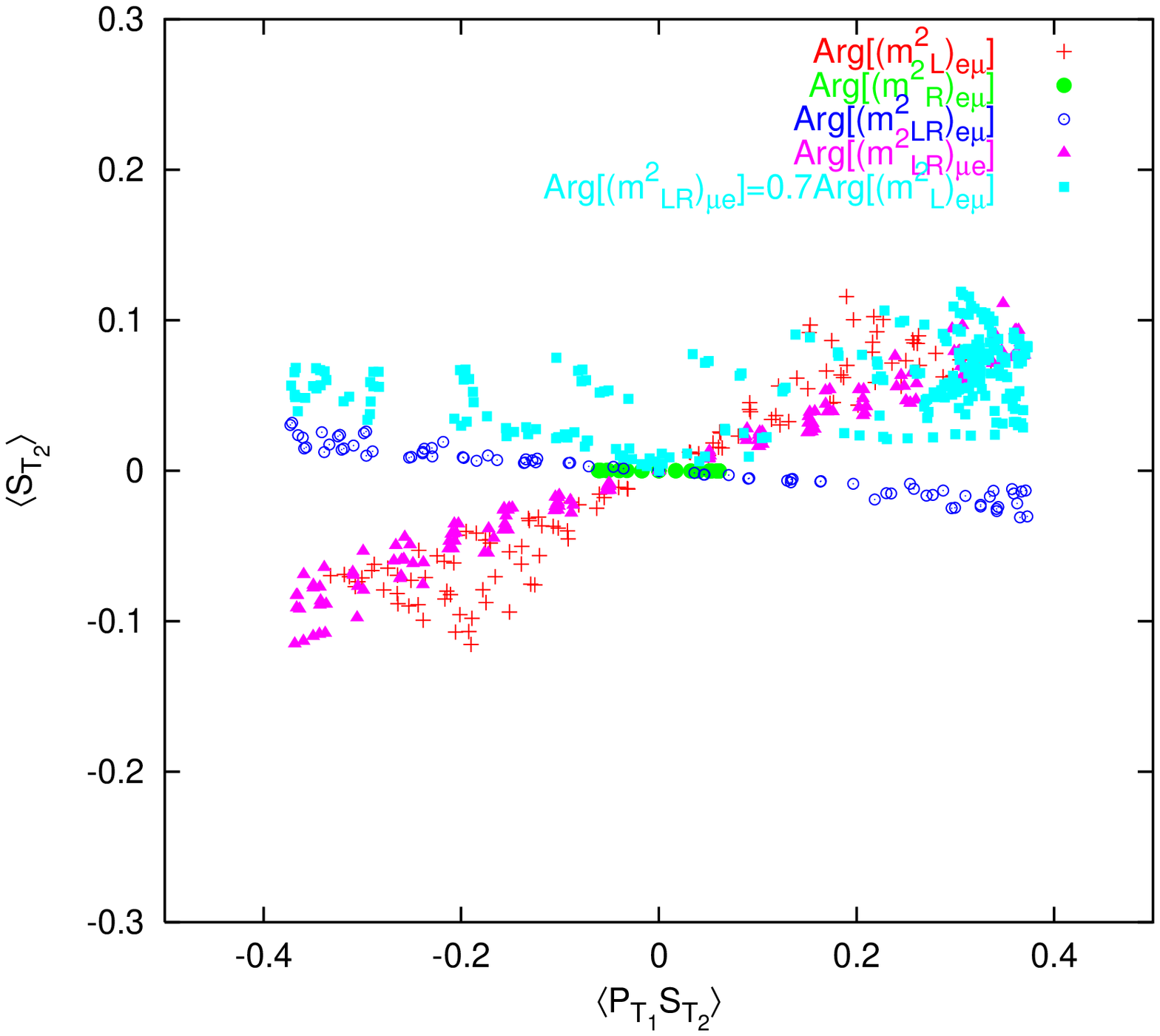}\includegraphics[scale=0.5]{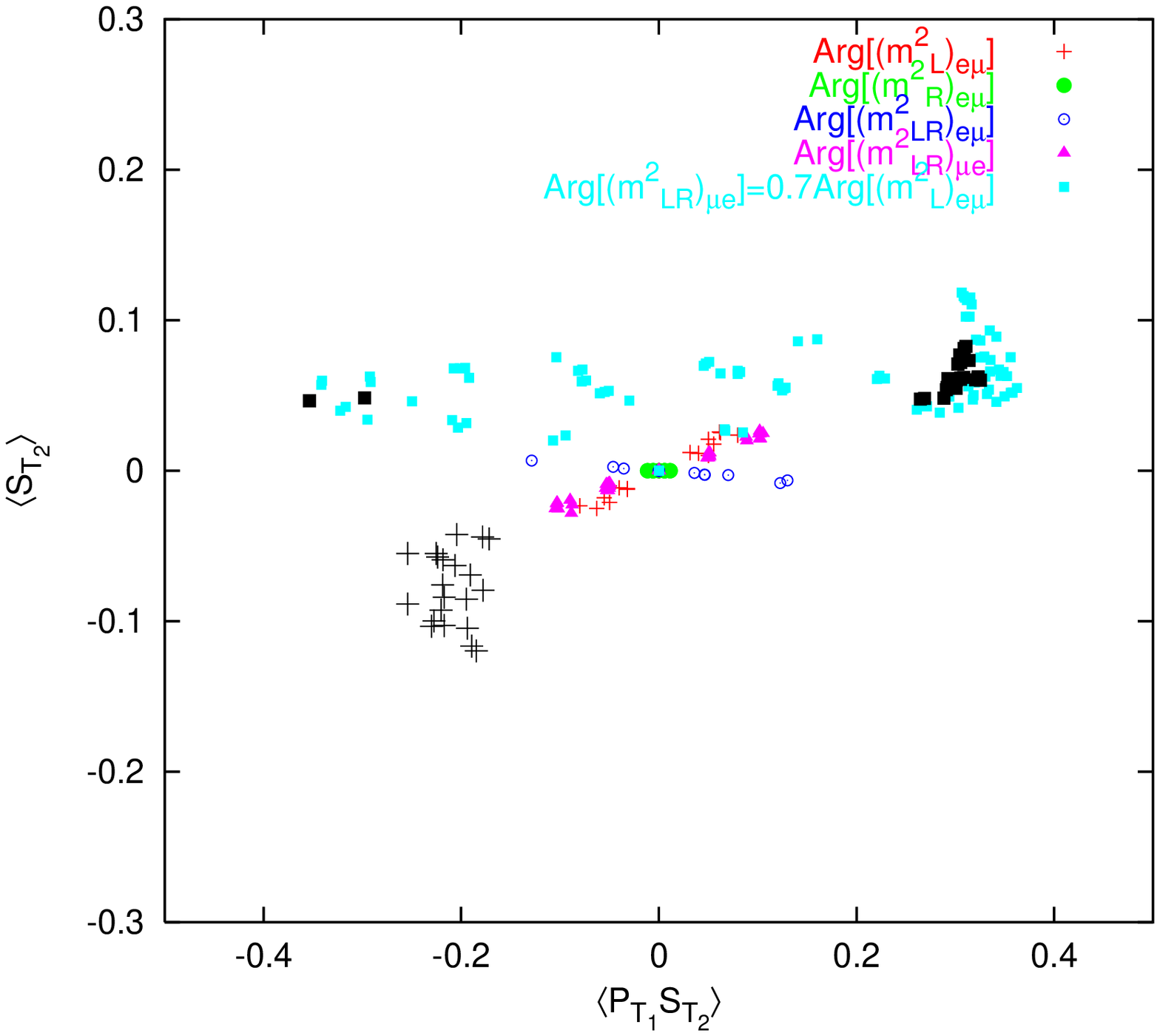}}
\centerline{\hspace{1cm}(a)\hspace{8cm}(b)}
\centerline{\vspace{-1.5cm}}\hspace{5mm}
\end{center}
\caption{ Transverse polarization in the $\mu \to e \gamma$ and
$\mu {\rm Ti} \to e {\rm Ti}$ processes. The input for LF
conserving parameters are the same as the input in
Fig.~\ref{sim-muegamma}: {\it i.e.,} the P3 benchmark with
$A_{\mu}=A_{e}$=700~GeV. The only sources of LFV are the $e\mu$
elements. In calculating $\overline{\langle P_{T_1}s_{T_2}
\rangle}$ (see Eq.~(\ref{average-theta2})) and $\overline{\langle
s_{T_2} \rangle}$ (see Eq.~(\ref{mean-conversion-sT2})) we have
respectively set $\mathbb{P}_\mu=100 \%$ and $\mathbb{P}_\mu=20
\%$. Points depicted by various colors and symbols as described in
the legend correspond to the case that the phases of various
elements vary between 0 and $2\pi$. The points show the
correlation of $\overline{\langle P_{T_1}s_{T_2} \rangle}$ and
$\overline{\langle s_{T_2} \rangle}$ at configurations of LFV for
which $0.3\leq$$R_1$$\leq0.4$, $0.7\leq$$R_2$$\leq0.9$,
$5.9\times10^{-12}$$\leq$$Br(\mu\rightarrow
e\gamma)$$\leq6.5\times10^{-12}$ and ${8.5\times10^{-14} \leq
R}(\mu {\rm Ti}\rightarrow e {\rm Ti})$$\leq1.1\times10^{-13}$. In
collecting the colored points in Fig.~(b) we have removed the
points for which $|d_e|$ exceeds $10^{-29}~e$~cm (the reach of
running experiments \cite{running}). The black points in Fig.~(b)
depicted by slightly larger plus and squares satisfy the condition
$2\times10^{-29}~e~\rm{ cm}<d_e< 3\times10^{-29}~e~{\rm cm}$. }
\label{degeneracy}
\end{figure}

Fortunately, the pairs of ($K_L,K_R$) and $(A_L,A_R)$ depend on
different combinations of the LFV elements. Thus, there is a hope
to solve a part of degeneracies by combining information from the
$(\mu \to e \gamma)$ and $(\mu N \to e N)$ experiments.
Fig.~\ref{degeneracy} demonstrates such a possibility. In the case
of the points depicted by red plus (+), green filled circle, dark
blue circle and purple triangle, all the phases are set to zero
except one of the phases which is specified in the legend and
varies between 0 and $2\pi$.  In the case of points depicted by
cyan squares, the phase of $ (m_{LR}^2)_{\mu e}$ is set equal to
0.7 of the phase of $(m_L^2)_{e\mu}$ which varies between zero and
$2\pi/0.7$ (thus, $\arg[ (m_{LR}^2)_{\mu e}]$ varies between zero
and $2\pi$).  The rest of the phases are set equal to zero. As we
saw in the previous section, the sensitivity to the phase of
$A_\mu$ is low (especially at the P3 benchmark) so in this
analysis we have not considered this phase and focused on the
effects of the phases of the LFV elements.

Hopefully, LHC will discover supersymmetry and  provide us with
information on the values of LF conserving parameters such as
values of $\tan \beta$ and the masses of neutralinos, charginos
(hence the values of $M_2$ and $\mu$) and sfermions and etc. In
the literature, it is discussed that under certain circumstances,
LHC can also measure the LFV parameters \cite{Diaz-Cruz}. However,
in this analysis, we solely rely on the LFV rare processes $\mu
\to e \gamma$ and $\mu N \to eN$ to derive the LFV parameters.
Having this prospect in mind, we have chosen the values at the P3
benchmark for the lepton flavor conserving parameters.
 We have then searched for  the values of the LFV $e \mu$
elements at which  the observable quantities Br$(\mu \to e
\gamma)$, $R(\mu {\rm Ti} \to e {\rm Ti})$, $R_1$ and $R_2$ are in
a given range. We have fixed $A_e$ and $A_\mu$ to 700~GeV. Notice
that measuring $|A_\mu|$ and $|A_e|$ at LHC is going to be
challenging if possible at all. In principle, we should have set
$A_\mu$ and $A_e$ as free parameters to be determined from the
$\mu \to e \gamma$ and $\mu-e$ conversion experiments along with
the LFV parameters. Notice however that, for $\tan \beta
\gtrsim10$, sensitivity to these parameters is low ({\it i.e.,}
varying $A_i$ from 0 to 700 GeV, the changes in the values of the
observables are less than 5\%). If $\tan \beta$ turns out to be
lower or a precision better than 5\% is achieved, $A_\mu$ and
$A_e$ should be treated as free parameters (rather than input).

The idea behind the plot is as follows. Suppose $\mu \to e \gamma$
and $\mu {\rm Ti} \to e {\rm Ti}$ are detected and their rates are
measured with some reasonable accuracy. Moreover suppose $R_1$ and
$R_2$ are measured and found to be in the range indicated in the
caption of Fig.~\ref{degeneracy}. The question is what
configurations of LFV elements and the CP-violating phases can
give rise to these values of the observables. To answer this
question, we have looked for the solutions by varying $|(m_L^2)_{e
\mu}|$, $|(m_R^2)_{e \mu}|$, $|(m_{LR}^2)_{e\mu}|$ and
$|(m_{LR}^2)_{\mu e}|$ respectively in the range ($0,10000)$$~{\rm
GeV}^2$, $(0,15000)$$~{\rm GeV}^2 $, ($0,50)$$~{\rm GeV}^2$ and
$(0,50)$$~{\rm GeV}^2$ for given values of the CP-violating
phases. We have then inserted the values of the LFV elements at
the solutions in the formulas of $\overline{\langle
s_{T_2}\rangle}$ and $\overline{\langle P_{T_1}s_{T_2}\rangle}$
and depicted it in Fig.~\ref{degeneracy}-a by a point.

From Fig.~\ref{degeneracy}-a, we observe that all sets of the
solutions depicted with various symbols reach to each other at the
point $\overline{\langle s_{T_2}\rangle}=\overline{\langle
P_{T_1}s_{T_2}\rangle}=0$. This is expected because setting the
phases equal to zero renders $A_L$, $A_R$, $K_L$ and $K_R$ real so
both $\overline{\langle s_{T_2}\rangle}$ and $\overline{\langle
P_{T_1} s_{T_2} \rangle}$ vanish ({\it see},
Eqs.~(\ref{average-theta2},\ref{mean-conversion-sT2})). Apart from
this point, the set of points depicted by plus and triangles  are
separate from points depicted by empty circles which means by
combining information from the $\mu \to e \gamma$ and $\mu -e$
conversion searches, one can solve the degeneracy between these
solutions. For example, if $|\overline{\langle
s_{T_2}\rangle}|<0.05$ and $\overline{\langle
P_{T_1}s_{T_2}\rangle}\simeq 0.38$, we can make sure that neither
of the solutions with zero $\arg[(m_{LR}^2)_{e\mu}]$ that we have
considered in this analysis can be the case. However, the
degeneracy  is not completely solved.  For example from
Fig.~\ref{degeneracy}-a, we observe that the regions over which
points depicted by plus and square are scattered, overlap. At the
intersection of the two regions, both ($\arg[(m_{LR}^2)_{\mu
e}]=0.7\arg[(m_L^2)_{e\mu}]\ne 0$) and ($\arg[(m_{LR}^2)_{\mu
e}]=0,~\arg[(m_L^2)_{e\mu}]\ne 0$) can be a solution.

We have repeated the same analysis for other ranges of $R_1$, $R_2$,
Br($\mu \to e \gamma$) and $R(\mu {\rm Ti} \to e {\rm Ti})$. As long
as $R_1$ and $R_2$ deviate from $\pm 1$, the above results are
maintained. However, when $R_1$ and $R_2$ approach $\pm 1$,
regardless of the values of the phases, the corresponding transverse
polarizations become so small that in practice cannot be measured.

 In summary, combining the information from $\mu \to e \gamma$
and $\mu N \to e N$ searches  considerably lifts the degeneracies
however, does not completely resolve them. By employing other
observables, it may be possible to completely solve the
degeneracies. For example,  it is in principle possible to derive
extra information on the $e\mu$ elements by studying other LFV
processes such as $\mu \to e \gamma\gamma$ which within our
scenario takes place with a rate suppressed by a factor of
$O(e^2/16 \pi^2)$ relative to the rate of $\mu \to e \gamma$. A
more promising approach is to employ the information from the
$d_e$ searches. As we discussed in the previous section, the
phases of the ${e\mu}$ elements can lead to $|d_e|\sim
10^{-29}~e$~cm which is within the reach of the currently running
experiments \cite{running}. To examine how much forthcoming
results on $d_e$ can help us to resolve the degeneracies, we have
presented Fig.~(\ref{degeneracy}-b). This figure is similar to
Fig.~(\ref{degeneracy}-a) with the difference that  at each point
in addition to observables in the $\mu \to e \gamma$ and $\mu-e$
conversion experiments, we have also calculated $d_e$. We have
removed the points for which $|d_e|>10^{-29}~e$~cm from the set of
points depicted by colored symbols. In the case of
$(\arg[(m_{LR}^2)_{e\mu}]=0,\arg[(m_L^2)_{e\mu}]\ne 0)$ and
$(\arg[(m_{LR}^2)_{e\mu}]=0.7\arg[(m_L^2)_{e\mu}]\ne 0)$, we have
also depicted points satisfying the condition $2\times
10^{-29}~e~{\rm cm} <d_e<3\times 10^{-29}~e~{\rm cm}$ with
slightly larger black symbols.

Notice that unlike in Fig. (a), in Fig.~(b) the regions over which
the squares and pluses are scattered have no overlap. This means
$d_e$ can help us to resolve the degeneracies. For example
according to Figs.~(\ref{degeneracy}-a,\ref{degeneracy}-b), if
$\langle s_{T_2}\rangle$ and $\langle P_{T_1}s_{T_2}\rangle$ are
measured and found to be respectively equal to 0.05 and 0.3, both
$(\arg[(m_{L}^2)_{e\mu}]=0,\arg[(m_{LR}^2)_{\mu e}]\ne 0)$ and
$(\arg[(m_{LR}^2)_{\mu e}]=0.7\arg[(m_L^2)_{e\mu}]\ne 0)$ can be a
solution. But if $d_e$ turns out to be in the range $(2-3)\times
10^{-29}~e~{\rm cm}$, the solution with
$(\arg[(m_{L}^2)_{e\mu}]=0)$ will be excluded.

%%%%%%%%%%%%%%%%%%%%%%%%%%%%%%%%%%%%%%%%%%%%%%%5555
%%%%%%%%%%%%%%%%%%%%%%%%%%%%%%%%%%%%%%%%%%%%%%%%5555
\section{Conclusions\label{results}}
%%%%%%%%%%%%%%%%%%%%%%%%%%%%%%%%%%%%%%%%%%%%%%%%%%55
%%%%%%%%%%%%%%%%%%%%%%%%%%%%%%%%%%%%%%%%%%%%%%%%5555555555
In this paper, we have first derived the formulas for the transverse
polarization of the final particles in  $\mu \to e \gamma$, $\mu \to
eee$ and $\mu-e$ conversion in terms of the couplings of the
effective LFV Lagrangian describing these processes. We have shown
that by measuring these polarizations, one can derive information on
the CP-violating phases of the underlying theory. We have then
focused on the polarizations of the final particles in the  $\mu \to
e \gamma$ and $\mu -e$ conversion processes. We have found that for
the configurations of LFV elements that asymmetries $R_1$ and $R_2$
(see Eqs.~(\ref{R1},\ref{R2})  for definitions) are not close to
$\pm 1$, the transverse polarization can be sizeable and sensitive
to certain combinations of the CP-violating phases. We therefore
suggest the following steps as the strategy to extract the
CP-violating phases. If in the future $\mu \to e \gamma$ and/or
$\mu-e$ conversion is detected with high statistics, it will be
possible to measure $R_1$ and/or $R_2$ by studying the angular
distribution of the final particles relative to the spin of the
decaying muon. If $R_1$ and/or $R_2$ turn out to considerably
deviate from $\pm 1$, it is then recommendable to equip the
experiment with polarimeters to measure the transverse polarizations
of the final particles and derive information on the phases of the
effective couplings.

The above results apply to a general beyond SM scenario that
provides large enough sources of LFV to allow detectable rates for
$\mu \to e \gamma$ and $\mu N \to eN$. Within a given scenario, the
couplings of the effective Lagrangian can depend on various
parameters in the underlying theory. This leads to degeneracies in
deriving these parameters. In this paper, we have addressed this
problem   in the context of $R$-parity conserving MSSM. We have
implicitly assumed that supersymmetry would be discovered at the LHC
and the lepton flavor conserving parameters relevant for this study
({\it e.g.,} chargino and neutralino masses, slepton and sfermion
masses and etc.) would be measured. We have then studied what can be
learnt about the LFV and CP-violating parameters of MSSM at $\mu \to
e\gamma$ and $\mu-e$ conversion experiments.

We have found that the dependence of the polarizations in the
cases of $\mu \to e \gamma$ and $\mu-e$ conversion  on the
parameters of the underlying theory is different. As a result,
depending on the configuration of the LFV elements, the effect can
be sizeable in none, only one or both of the $\mu \to e\gamma$ and
$\mu-e$ conversion processes. Thus, the polarization studies in
these processes are complementary.

We have focused on the effect of the $e\mu$ elements and studied
the dependence of the various observables on the  phases of
$A_\mu$ and the $e\mu$ LFV elements.  Since there are already
strong bounds on the phases of $\mu$, $M_1$ (Bino mass) and $A_e$
from electric dipole moment searches, we have taken these
parameters real. We have found that for most parts of the
parameter space with large $\tan \beta$ ({\it i.e.,}
$\tan\beta\sim 10$) the sensitivity to $A_\mu$ is low but the
sensitivity  of transverse polarizations both in $\mu \to e
\gamma$ and $\mu-e$ conversion to $\arg[(m_L^2)_{e\mu}]$ is high.
However, there are regions in the parameter space that the
sensitivity to $\arg[A_\mu]$ is sizeable ({\it e.g.,} the $\delta$
benchmark \cite{delta}). The sensitivity to $\arg[(m_R^2)_{e\mu}]$
in the case of $\mu \to e\gamma$ is also high but in the case of
the $\mu-e$ conversion, the sensitivity to $\arg[(m_R^2)_{e\mu}]$
is low.

In the context of the present scenario, various CP-violating
parameters can affect the observables in the $\mu \to e\gamma$ and
$\mu N\to e N$ experiments. These polarizations also strongly
depend on the ratios of the absolute values of the various LFV
elements. We have shown that for configurations of LFV elements
for which $-0.9<R_1,R_2<0.9$, combining information on $R_1$,
$R_2$, Br$(\mu \to e \gamma)$ and $R(\mu N \to e N)$ with
information on the transverse polarization of the final particles
can help us to considerably decrease degeneracies and derive
information on these phases. However, information from these
measurements is not enough to fully resolve degeneracies. For
example, we have shown degeneracies between solutions
$(\arg[(m_{LR}^2)_{\mu e}]=0,\arg[(m_L^2)_{e\mu}]\ne 0)$ and
$(\arg[(m_{LR}^2)_{\mu e}]=0.7\arg[(m_L^2)_{e\mu}]\ne 0)$ cannot
be removed even when we use all the information accessible at the
$\mu\to e \gamma$ and $\mu N \to e N$ search experiments. To fully
resolve the degeneracies, extra information from other experiments
has to be employed. We have also demonstrated that the forthcoming
results of the $d_e$ search can help us to remove the degeneracies
further.

 Notice that by [simultaneously] turning on the $\mu \tau$ and
$e\tau$ elements, more degeneracies will emerge. To resolve these
degeneracies, one can employ other observables such as Br($\tau
\to e \gamma$) and Br($\tau \to \mu \gamma$). Studying the general
case is beyond the scope of the present paper and will be
presented elsewhere.

We have also briefly discussed the possibility to derive further
information by using different nuclei in the $\mu -e$ conversion
experiment and found that since the ratio of proton number to the
neutron number for different nuclei is close to each other, the
polarizations are similar for different nuclei. Unless a precision
better than 5\% is achieved, changing the nuclei will not help us
to extract information on an extra combination of the parameters
but can be considered as a cross-check of the results.

\section*{Acknowledgement} We would like to thank M. M.
Sheikh-Jabbari for careful reading of the manuscript and the useful
remarks.

\end{document}